\documentclass[11pt]{article}
\usepackage[utf8x]{inputenc}
\usepackage{mathtools}
\usepackage{amsfonts}
\usepackage{amsmath}
\usepackage{amssymb}
\usepackage{dsfont}
\usepackage{slashed}
\usepackage{fullpage}
\usepackage{cite}
\usepackage[colorlinks=true,linkcolor=blue,citecolor=magenta,linktocpage=true]{hyperref}
\usepackage{color}
\usepackage{titlesec}
\usepackage{braket}
\usepackage{caption}
\usepackage{float}
\titleformat*{\section}{\normalsize\bfseries}
\titleformat*{\subsection}{\normalsize\bfseries}
\titleformat*{\subsubsection}{\normalsize\bfseries}
\makeatletter
\renewcommand{\@dotsep}{1000}

\definecolor{MyDarkRed}{rgb}{0.7,0,0}

\definecolor{MyBlue}{rgb}{0.0,0.0,.5}

\def\be#1\ee{\begin{align}#1\end{align}}
\def\ba{\begin{eqnarray}}
\def\ea{\end{eqnarray}}
\def\q{\qquad}
\def\f{\frac}
\def\sgn{\mathrm{sgn}}
\def\eps{\varepsilon}

\def\de{\mathrm{d}}

\def\mb{\bar{\mu}}
\def\lp{\ell_\text{Pl}}

\def\B{\mathcal{B}}
\def\C{\mathcal{C}}

\def\H{\mathcal{H}}

\def\N{\mathcal{N}}

\def\R{\mathbb{R}}
\def\S{\mathcal{S}}
\def\T{\mathcal{T}}
\def\U{\mathcal{U}}

\numberwithin{equation}{section}

\begin{document}

\title{\Large{\textbf{\sffamily Quantum dynamics of the black hole interior in LQC}}}
\author{\sffamily Francesco Sartini, Marc Geiller}
\date{\small{\textit{
Univ Lyon, ENS de Lyon, Univ Claude Bernard Lyon 1,\\ CNRS, Laboratoire de Physique, UMR 5672, Lyon, France}}}

\maketitle

\begin{abstract}
It has been suggested that the homogeneous black hole interior spacetime, when quantized following the techniques of loop quantum cosmology, has a resolved singularity replaced by a black-to-white hole transition. This result has however been derived so far only using effective classical evolution equations, and depends on details of the so-called polymerization scheme for the Hamiltonian constraint. Here we propose to use the unimodular formulation of general relativity to study the full quantum dynamics of this mini-superspace model. When applied to such cosmological models, unimodular gravity has the advantage of trivializing the problem of time by providing a true Hamiltonian which follows a Schr\"odinger evolution equation. By choosing variables adapted to this setup, we show how to write semi-classical states agreeing with that of the Wheeler--DeWitt theory at late times, and how in loop quantum cosmology they evolve through the would-be singularity while remaining sharply peaked. This provides a very simple setup for the study of the full quantum dynamics of these models, which can hopefully serve to tame regularization ambiguities.
\end{abstract}

\thispagestyle{empty}
\newpage
\setcounter{page}{1}

\hrule
\tableofcontents
\vspace{0.7cm}
\hrule


\newpage

\section{Introduction}

Quantum gravity is expected to shed light onto the fate of the singularities which appear in classical general relativity. The most notable occurrences of these are in cosmological models, e.g. the big-bang in FLRW-like spacetimes, and inside of black holes. Understanding how this classical singular behavior is affected by quantum theory is therefore fundamental in order to get a complete description of the evolution of our Universe, and to know what becomes of black holes after they have evaporated. With the advent of precision cosmology \cite{Martin:2003bt,Primack:2004gb,Ade:2013sjv,Martin:2013tda,Abbott:2018lct} and the recent observational access to black hole physics \cite{Abbott:2016blz,Akiyama:2019cqa}, the prospect of testing models of quantum gravity is also becoming more realistic \cite{Agullo:2012sh,Barrau:2015uca,Ashtekar:2016wpi,Rovelli:2017zoa,Bianchi:2018mml,Giddings:2019jwy,Barrau:2020ixa,Ashtekar:2020gec,Ashtekar:2020ifw}. In particular, models in which the (big-bang or black hole) singularity is replaced by a bounce open a new window onto phenomenology.

Lots of work has been devoted to the study of bounces driven by quantum effects within loop quantum gravity \cite{Thiemann:2001yy,Ashtekar:2004eh} and inspired symmetry-reduced models of quantum cosmology \cite{Bojowald_2015}. In loop quantum cosmology (LQC hereafter) \cite{Ashtekar:2003hd,Ashtekar_2011}, there is robust evidence that the big-bang singularity is replaced by a quantum bounce \cite{Bojowald:2001xe,Ashtekar:2006uz,Ashtekar:2006wn}. Tentatives to extend these FLRW results to black hole spacetimes have been numerous. In particular, one can distinguish approaches which are based on symmetry-reduced models \cite{Ashtekar:2005qt,Modesto:2005zm,Modesto:2006mx,Bohmer:2007wi,Campiglia:2007pb,Brannlund:2008iw,Chiou:2008eg,Chiou:2008nm,Corichi:2015xia,Dadhich:2015ora,Saini:2016vgo,Olmedo:2017lvt,Cortez:2017alh,Ashtekar:2018cay,Ashtekar:2018lag,Bodendorfer:2019xbp,Bodendorfer:2019cyv,Bodendorfer:2019nvy,Bodendorfer:2019jay,Ashtekar:2020ckv}, approaches within the full theory \cite{Alesci:2018loi,Alesci:2019pbs,Alesci:2020zfi}, and other bouncing models with various phenomenological inputs (not necessarily coming from loop quantum gravity) \cite{Haggard:2015iya,BenAchour:2017ivq,BenAchour:2020gon,BenAchour:2020bdt,BenAchour:2020mgu,DAmbrosio:2020mut,Martin-Dussaud:2019wqc,Rovelli:2018cbg}. In all these models, the black hole singularity is replaced by a non-singular phase where curvature remains finite, and beyond which one may find a white hole or a deSitter universe \cite{Alesci:2018loi,Alesci:2019pbs,Alesci:2020zfi}.

The difficulty with models applying the techniques of LQC to the black hole interior is that they rely on classical effective evolution equations. In this effective approach, one is using a classical modification of the Hamiltonian constraint, via the so-called polymerization of a preferred choice of phase space variables, in order to derive effective evolution equations. While in LQC applied to FLRW models it has been shown that the effective equations approximate well the numerical evolution of quantum states which are semi-classical in the future \cite{Taveras:2008ke}, such an analysis, and that of the quantum dynamics in general, is still missing in the context of black holes.

Reliance on the effective equations poses another challenge, which is to define a consistent regularization of the Hamiltonian constraint. In the absence of guidance from the full theory, the heuristic regularization scheme which is adopted in effective approaches is to replace a choice of phase space variables (say) $q$ by their ``polymerized'' version $\sin(\lambda q)/\lambda$, where $\lambda$ is an ultraviolet cutoff typically related to the area gap of full LQG, and which is furthermore allowed to be phase space dependent. This compactification of variables is expected to capture the effects of quantum geometry, imported in mini-superspace from the full theory. Evidently, this procedure requires to decide on a choice of phase space variables to polymerize, and on the functional form of the regulator $\lambda$. In FLRW models, there exist solid arguments in favor of a unique choice known as the $\bar{\mu}$-scheme (although additional ambiguities remain, see e.g. \cite{Perez:2005fn,Vandersloot:2005kh,BenAchour:2016ajk,Bojowald:2020wuc}). For the black hole interior however, there is no agreement on the choice of variables and polymerization scheme, and this has recently been the subject of many discussions \cite{Ashtekar:2018lag,Ashtekar:2018cay,Bodendorfer:2019xbp,Bodendorfer:2019cyv,Bodendorfer:2019nvy,Bodendorfer:2019jay,Ashtekar:2020ckv}. The effective dynamics approach, in which one uses heuristic deformations of the constraints, also raises the important question of whether the models so-constructed posses spacetime covariance \cite{Bojowald:2015zha,BenAchour:2018khr,Bojowald:2019dry,Bojowald:2020xlw,Bojowald:2020unm,Bojowald:2020dkb}, which is another important consistency check on the regularized constraints\footnote{Although for this one needs to go beyond homogeneity in order to have a non-trivial spacetime algebra of constraints.}. While the homogeneous model does not allow to discuss a non-trivial algebra of constraints, it actually has an $\mathfrak{iso}(2,1)$ Poincar\'e symmetry, as revealed and studied in \cite{CVH-BH}, which can be used as a guide towards a symmetry-preserving regularization.

Once a regularization scheme has been adopted, one can ask the question of the quantum dynamics, and then study if the effective classical equations indeed emerge from this quantum theory. Whether one is studying Wheeler--DeWitt (WDW) mini-superspace quantization, or quantization on the polymer Hilbert space inherited from LQG, the study of the quantum dynamics requires to deal with the usual difficulties present in general relativity, namely the problem of time and that of finding observables \cite{Isham:1992ms,Hoehn:2019owq}. In the study of LQC for FLRW models, analytical results are typically derived by deparametrizing the theory with respect to a scalar field clock, thereby interpreting the Hamiltonian constraint as an evolution operator with respect to an internal time. The positive and negative frequency solutions to the (initially) quadratic evolution equation in scalar field time then satisfy a Schr\"odinger evolution equation which is formally the square root of the Klein--Gordon equation. Evidently, this procedure may not be well-defined for more complicated gravitational Hamiltonians, or for arbitrary matter fields (e.g. a scalar field with a potential).

A great simplification occurs however when considering unimodular gravity \cite{PhysRevD.40.1048,unimodular_web}. This is a formulation of gravity equivalent to usual general relativity, but which allows for the cosmological constant to vary between different solutions (it appears as a simple integration constant). In the canonical theory, the lapse is fixed to a particular value and the theory possesses a true Hamiltonian which generates evolution along a ``cosmological clock'' variable canonically conjugated to the cosmological constant. When applied to mini-superspace models, unimodular gravity completely trivializes the problem of time, and gives rise to a Schr\"odinger evolution equation with respect to which unitary can be unambiguously defined. This has motivated recent studies of the information loss problem within FLRW cosmological models, where the possibility of transferring information to Planckian degrees of freedom (such as those of LQG) has been proposed as a mechanism behind the apparent non-unitarity observed by low energy coarse grained observers \cite{Amadei:2019ssp,Amadei:2019wjp}.

Unimodular gravity was applied to the LQC quantization of FLRW models in \cite{Chiou:2010ne} (see also \cite{Riahi:2018gvm}). The unimodular representation becomes however truly useful in more complicated models, and for this reason we propose to apply it to the homogeneous black hole interior spacetime. The goal of this work is to advocate for the use of the unimodular clock variable in the study of quantum cosmological models, both within traditional WDW and LQC quantization. In the context of black hole interior models, the motivation is, on the one hand, to eventually extend scenarios such as the one proposed in \cite{Amadei:2019ssp,Amadei:2019wjp} in order to study the information loss problem. On the other hand, the simple quantum theory resulting from the use of the unimodular clock could be used in order to test the various regularization schemes which have been proposed in the literature (and their related choices of canonical variables). Starting from a preferred choice which we will argue in favor of, we provide a proof of principle construction which can be extended to other proposals in order to test their viability and the properties of the resulting quantum theory. Note that \cite{Bodendorfer:2019cyv} has given ingredients of the quantum theory and in particular found the kernel of the Hamiltonian constraint operator (with a given choice of variables and polymerization). Here we aim at constructing instead the actual quantum evolution of states by solving a deparametrized evolution equation.

This article is organized as follows. We first recall in section \ref{sec:2} the classical structure of the black hole interior with cosmological constant. In section \ref{sec:effective} we review the effective classical dynamics arising from various polymerization schemes. We introduce a ``mixed scheme'' in \ref{sec:mixed scheme}, based on the variables introduced in \cite{Bodendorfer:2019nvy,Bodendorfer:2019cyv}. This allows to simplify the analysis of the quantum theory by finding simple eigenfunctions for the WDW operator, which can then be used to compute the evolution of LQC states across the singularity. In both the classical and effective evolutions, we discuss the role of the Dirac observables. Finally, we present in section \ref{sec:quantum} the construction of the quantum dynamics of the theory.

\section{Black hole interior with cosmological constant}
\label{sec:2}

Let us start by studying the classical setup underlying our construction. We will present the metric for the black hole interior with cosmological constant, the corresponding LQG connection and triad variables, and the classical Hamiltonian. We will then use this Hamiltonian to compute and study the classical equations of motion in terms of the cosmological time variable.

\subsection{Variables and Hamiltonian}

Because we plan on using the cosmological time variable to construct the classical and quantum evolutions, we cannot restrict ourselves to the study of the standard Schwarzschild black hole interior. Instead, we need to consider the more general Schwarzschild--de Sitter (SdS) space-time. In static coordinates, the SdS line element takes the form
\be\label{SdS}
\de s^2=-f(r)\de\tilde{t}^{\,2}+f(r)^{-1}\de r^2+r^2\de\Omega^2,
\ee
where $\de\Omega^2=\de\theta^2+\sin^2\theta\,\de\phi^2$ is the metric on the unit 2-spheres at constant $r$ and $\tilde{t}$. The function $f(r)$ is given by
\be
f(r)\coloneqq1-\f{2M}{r}-\f{\Lambda}{3}r^2,
\ee
where $M$ is the mass of the black hole and $\Lambda$ is the cosmological constant (and we have set $G=1=c$). To get a preliminary understanding of the geometry of this space-time, it is useful to analyse the positive real roots of the cubic equation $f(r)=0$.
\begin{itemize}
\item For $\Lambda>0$ and $9M^2\Lambda<1$, there are three real roots and two are positive, corresponding respectively to a black hole horizon at 
\be\label{horizon radius}
\f{r_\text{h}}{M}=-\f{2}{M\sqrt{\Lambda}}\cos\left(\f{1}{3}\arccos\left(3M\sqrt{\Lambda}\right)-\f{2\pi}{3}\right)=2+\f{8}{3}M^2\Lambda+\mathcal{O}(M\sqrt{\Lambda})^4,
\ee
and a cosmological horizon at
\be
\f{r_\text{c}}{M}=-\f{2}{M\sqrt{\Lambda}}\cos\left(\f{1}{3}\arccos\left(3M\sqrt{\Lambda}\right)+\f{2\pi}{3}\right)=\f{\sqrt{3}}{M\sqrt{\Lambda}}-1+\mathcal{O}(M\sqrt{\Lambda}).
\ee
\item For $\Lambda>0$ and $9M^2\Lambda=1$, there are three real roots and two are degenerate and positive, corresponding to a single horizon at $r_\text{h}=3M$.
\item For $\Lambda>0$ and $9M^2\Lambda>1$ there are two imaginary roots and a negative real root, so no horizons.
\item For $\Lambda<0$, there are two imaginary and a positive real root, corresponding to an horizon at
\be
\f{r_\text{h}}{M}=-\f{2}{M\sqrt{-\Lambda}}\sinh\left(\f{1}{3}\,\text{arcsinh}\left(3M\sqrt{-\Lambda}\right)\right)=2+\f{8}{3}M^2\Lambda+\mathcal{O}(M\sqrt{-\Lambda})^4.
\ee
\end{itemize}
In the rest of this work we are going to consider the case $\Lambda>0$. Notice that, in principle, the condition $9M^2\Lambda<1$ gives an upper limit on the mass of black holes, although with the observed value of the cosmological constant this bound is very loose. For example, for a SdS black hole of solar mass, the radius of the cosmological horizon is of the order of the Hubble radius.

Just like in the case $\Lambda=0$, the SdS metric \eqref{SdS} admits a maximal analytic extension. More importantly for our purposes, it is also possible to find homogeneous Kantowski--Sachs coordinates covering the region located inside of the black hole horizon, and with which the line element becomes
\be\label{KdS}
\de s^2
&=f(t)^{-1}\de t^2-f(t)\de x^2+t^2\de\Omega^2\cr
&=-\left(\f{2M}{t}+\f{\Lambda}{3}t^2-1\right)^{-1}\de t^2+\left(\f{2M}{t}+\f{\Lambda}{3}t^2-1\right)\de x^2+t^2\de\Omega^2.
\ee
In these homogeneous coordinates the spatial slices are $\Sigma=\mathbb{R}\times\mathbb{S}^2$ with $x\in\mathbb{R}$, the singularity is located at $t=0$, and the horizon is at $t=r_\text{h}$. Since this metric is homogeneous we can apply to it the quantization techniques of LQC.

For this, we need to set up the Hamiltonian formulation using canonically conjugated connection and triad variables. The details of this construction are given in appendix \ref{appendix:EAH}. Following the abundant literature on Kantowski--Sachs LQC, we consider the line elements of the form
\be\label{homogeneous metric}
\de s^2=-N^2\de t^2+\f{p_b^2}{L_0^2|p_c|}\de x^2+|p_c|\de\Omega^2,
\ee
with $N(t)$, $p_b(t)$ and $p_c(t)$ the three time-dependent functions, and where $L_0$ is a fiducial length parameter used to regulate the non-compact spatial integrations in the Hamiltonian and symplectic structure. This metric is singular both for $p_b=0$ and $p_c=0$. However, only $p_c=0$ represents a true singularity (it is possible to show that the Kretschmann scalar behaves as $p_c^{-1}$), while $p_b=0$ is simply an horizon singularity.

With the form \eqref{homogeneous metric} of the metric, one can show that the densitized triad components $E^a_i$ and the canonically conjugated connection components $A^i_a$ take the form\footnote{From this one can see that only the ratios $L_0^{-1}p_b$ and $L_0^{-1}c$ have an invariant meaning under the rescaling $L_0\to\alpha L_0$ of the fiducial length.}
\begin{subequations}
\be
E^a_i\tau^i\partial_a&=p_c\sin\theta\,\tau^1\partial_x+\f{p_b}{L_0}\sin\theta\,\tau^2\partial_\theta+\f{p_b}{L_0}\tau^3\partial_\phi,\\
A_a^i\tau_i\de x^a&=\f{c}{L_0}\tau_1\de x+b\tau_2\de\theta+b\sin\theta\,\tau_3\de\phi-\cos\theta\,\tau_1\de\phi.
\ee
\end{subequations}
The function $N$ is analogous to the lapse of canonical gravity, and as such is pure gauge. We are therefore considering a symmetry-reduced phase space spanned by the two canonical pairs $(b,p_b)$ and $(c,p_c)$. One can see that this classical parametrization of the line element has the orientation reversal symmetry $(p_b,p_c)\to-(p_b,p_c)$, which we will fix by restricting ourselves to $(p_b,p_c)\geq0$.

The classical dynamics of the theory is defined by the Hamiltonian and the Poisson bracket relations between the canonical phase space variables. This construction is recalled in appendix \ref{appendix:EAH}. The Hamiltonian is
\be\label{Hamiltonian}
\H=-\f{N}{2\gamma^2\sqrt{p_c}}\Big(2bcp_c+(b^2+\gamma^2)p_b\Big)+\f{N\Lambda}{2} p_b\sqrt{p_c}\approx0,
\ee
where we have included a cosmological constant matter term, as required in order to work with the unimodular time variable canonically conjugated to $\Lambda$, and where $\gamma$ is the Barbero--Immirzi parameter\footnote{In all numerical calculations we use the value $\gamma=0.2375$.}. The weak equality denotes the fact that the Hamiltonian is vanishing as a constraint. The canonical Poisson brackets are
\be\label{PBs}
\{b,p_b\}=\gamma,
\q\q
\{c,p_c\}=2\gamma,
\q\q
\{T,\Lambda\}=8\pi.
\ee
The first two brackets concern the gravitational degrees of freedom of the theory, while the last one encodes the fact that in unimodular gravity, as recalled in appendix \ref{appendix:unimodular}, the cosmological constant $\Lambda$ is canonically conjugated to the cosmological time variable $T$.

\subsection{Classical dynamics}
\label{sec:classical}

With the classical Hamiltonian at our disposal, we can compute and solve the equations of motion of the system. Since the idea of the present work is to use the cosmological time variable and the corresponding choice of lapse in order to deparametrize the time evolution in the quantum theory, it is natural to also proceed to the same deparametrization at the classical level. However, we defer this calculation to appendix \ref{appendix:unimodular N}, and choose here instead the usual lapse
\be\label{standard N}
N=\f{\gamma\sqrt{p_c}}{b}
\ee
in order to facilitate comparison with the existing literature. With this choice of lapse the Hamiltonian is
\be\label{H with standard N}
\H=-\f{1}{2\gamma b}\Big(2bcp_c+(b^2+\gamma^2)p_b\Big)+\f{\gamma\Lambda}{2}\f{p_bp_c}{b}.
\ee

The time evolution of a phase space function $g$ is denoted by a dot and given by the Poisson bracket $\dot{g}=\{g,\H\}$. For the phase space variables of interest we find
\begin{subequations}\label{EOMs with standard N}
\be
\dot{b}&=\f{1}{2b}\Big(\gamma^2\Lambda p_c-(b^2+\gamma^2)\Big),\\
\dot{c}&=\gamma^2\Lambda\f{p_b}{b}-2c,\\
\dot{p}_b&=\f{p_b}{2b^2}\Big(\gamma^2\Lambda p_c+(b^2-\gamma^2)\Big),\\
\dot{p}_c&=2 p_c,\\
\dot{T}&=4 \pi \gamma \f{p_b p_c}{b},\\
\dot{\Lambda}&=0.
\ee
\end{subequations}
In addition to these dynamical equations we have the relation
\be\label{c on constraint surface}
c\approx\f{p_b}{2bp_c}\Big(\gamma^2\Lambda p_c-(b^2+\gamma^2)\Big),
\ee
which follows from the vanishing of the Hamiltonian constraint. The equation of motion on $\Lambda$ simply states that this variable is indeed constant. Fortunately, the presence of $\Lambda\neq0$ still allows to find the solutions to the equations of motion in closed form. They are given by
\begin{subequations}\label{classical solutions with standard N}
\be
b(\tau)&=\pm\gamma\left(e^{-(\tau-\tau_0)}+\f{\Lambda}{3}Be^{2(\tau-\tau_0)}-1\right)^{1/2},\\
c(\tau)&=\gamma A\left(\f{\Lambda}{3}e^{(\tau-\tau_0)}-\f{1}{2B}e^{-2(\tau-\tau_0)}\right),\\
p_b(\tau)&=\pm Ae^{(\tau-\tau_0)}\left(e^{-(\tau-\tau_0)}+\f{\Lambda}{3}Be^{2(\tau-\tau_0)}-1\right)^{1/2},\label{pb classical solution}\\
p_c(\tau)&=Be^{2(\tau-\tau_0)},\\
T(\tau)&=\pm\f{4\pi}{3}ABe^{3(\tau-\tau_0)}-T_0,
\ee
\end{subequations}
where $(\tau_0,T_0,A,B)$ are integration constants. The integration constants $\tau_0$ and $T_0$ can be set to zero without loss of generality, as they reflect the gauge freedoms in shifting respectively the time coordinate $\tau$ and the cosmological time $T$. The two other constants can be rewritten as $A=4D/3$ and $B=CD^2/9$ in terms of first integrals of the motion given by\footnote{We have chosen the constant numerical factors in the definition of $C$ and $D$ in such a way that their expressions in terms of the variables used for the quantization later on in section \ref{sec:quantum} will be simpler.}
\be\label{Dirac observables CD}
C=\f{16}{\gamma^2}\f{b^2p_c}{p_b^2},
\q\q
D=\f{1}{2\gamma}(bp_b-cp_c)+\f{\gamma}{2}\f{p_b}{b}.
\ee
One can readily verify that $\{C,\H\}=0=\{D,\H\}$, so in the terminology of constrained Hamiltonian systems these are the two Dirac observables of the theory.

With the solutions to the classical equations of motion at hand, one can go back to the parametrization \eqref{homogeneous metric}, and write this line element as
\be
\de s^2=-Be^{2\tau}\left(e^{-\tau}+\f{\Lambda}{3}Be^{2\tau}-1\right)^{-1}\de\tau^2+\f{A^2}{L_0^2B}\left(e^{-\tau}+\f{\Lambda}{3}Be^{2\tau}-1\right)\de y^2+Be^{2\tau}\de\Omega^2.
\ee
Introducing the new coordinates
\be\label{new coordinates}
t=\sqrt{B}e^\tau,
\q\q
x=\f{A}{L_0\sqrt{B}}y,
\ee
one of the integration constants gets reabsorbed, and we find the homogeneous interior line element \eqref{KdS} with a mass
\be
M=\f{\sqrt{B}}{2}=\f{D\sqrt{C}}{6}.
\ee
Note that we can rewrite the Dirac observables as
\be
2M=\f{D\sqrt{C}}{3}\approx\f{\sqrt{p_c}}{\gamma^2}(b^2+\gamma^2)-\f{\Lambda}{3}p_c^{3/2},
\q\q
-2\gamma D\approx3cp_c-\gamma^2\Lambda\f{p_bp_c}{b},
\ee
where for the weak equality we have used the constraint \eqref{c on constraint surface}. In the case $\Lambda=0$ these reduce to the Dirac observables identified in \cite{Bodendorfer:2019jay}, and $2M$ then corresponds to the radius of the horizon as can be seen from \eqref{horizon radius}.

\section{Effective dynamics}
\label{sec:effective}

We now turn to the study of the effective classical dynamics. The idea behind this construction is to modify the classical Hamiltonian using a so-called polymerization scheme, supposed to encode semi-classical corrections to the classical dynamics discussed in the previous section. Inspired by the full theory, where the connection is not available as an operator on the chosen Hilbert space and has to be replaced by holonomies, this polymerization is typically implemented by replacing the connection variables $(b,c)$ in the classical Hamiltonian by\footnote{The fact that only $L_0^{-1}c$ has invariant meaning under rescaling of the fiducial length $L_0$ constrains the associated polymerization parameter to be such that only $L_0\delta_c$ has an invariant meaning.}
\be
\S(b)\coloneqq\f{\sin(b\delta_b)}{\delta_b},
\q\q
\S(c)\coloneqq\f{\sin(c\delta_c)}{\delta_c}.
\ee
The precise form and phase space dependency of the regulators $(\delta_b,\delta_c)$ has been the subject of much debate in the literature in recent years. A further freedom in this construction is that of choosing the canonical variables themselves, and of considering polymerization schemes for variables which do not necessarily correspond to the components of the connection.

In flat FLRW LQC, the so-called improved $\mb$ scheme provides a viable and much studied prescription for choosing the regulator (of the unique homogeneous connection variable). Since no such robust choice is available in the case of the black hole interior, we would like to take here a first step towards the construction of a deparametrized quantum theory, which one could then use as a testbed for various proposals. This construction is however a bit circular, as we need to base our quantum theory on a choice of canonical variables and polymerization scheme. We will do this by adapting the $(p,v)$-type variables of \cite{Bodendorfer:2019nvy,Bodendorfer:2019cyv}, on which we will implement a ``mixed'' polymerization scheme adapted to the study of the quantum theory with unimodular clock variable. In order to understand this construction, presented in section \ref{sec:mixed scheme}, we will first very briefly review the main choices of polymerization existing in the literature. We refer the reader to \cite{Ashtekar:2018cay,Bodendorfer:2019jay} for very detailed reviews of the various schemes, with in particular an in-depth study of the role of the (classical and effective) Dirac observables in \cite{Bodendorfer:2019jay}.

\subsection[$\bar{\mu}$ scheme]{$\boldsymbol{\bar{\mu}}$ scheme}
\label{sec:effective mubar}

As done in \cite{Bohmer:2007wi}, it is natural to start by trying to adapt to the black hole interior case the improved $\bar{\mu}$ scheme initially developped in FLRW LQC. In this scheme, the polymerization parameters $\delta_b$ and $\delta_c$ for the holonomies depend on the triad variables $p_b$ and $p_c$. To define the exact dependency, the physical input is to constrain the physical area enclosed by holonomy loops to be equal to the area gap of LQG, namely $\Delta=2\sqrt{3}\pi\gamma\lp^2$. The classical physical area of a loop in the $(x,\theta)$ plane is given by $\text{Ar}_{x\theta}=\delta_b\delta_cp_b$. On the other hand, the loop on the 2-sphere $(\theta,\phi)$ is not closed. However, as discussed in \cite{Bohmer:2007wi}, because of homogeneity it is possible to assign to it an effective area $\text{Ar}_{\theta\phi}=\delta_b^2p_c$. Imposing that these two areas be equal to $\Delta$ and solving for the $\delta$'s leads to
\be
\delta_b=\f{\sqrt{\Delta}}{\sqrt{p_c}},
\q\q
\delta_c=\f{\sqrt{\Delta p_c}}{p_b}.
\ee
Note that this is compatible with the invariance under choice of $L_0$.

With the choice of lapse \eqref{standard N}, replacing the connection variables $(b,c)$ by their polymerized counterpart in \eqref{H with standard N} leads to the effective Hamiltonian constraint
\be\label{effective Hamiltonian}
\H_\text{eff}=-\f{1}{2\gamma\S(b)}\Big(2\S(b)\S(c)p_c+\big(\S(b)^2+\gamma^2\big)p_b\Big)+\f{\gamma\Lambda}{2}\f{p_bp_c}{\S(b)}.
\ee
Taking into account the fact that the polymerization parameters $(\delta_b,\delta_c)$ are phase space functions, the effective equations of motion take the form
\begin{subequations}\label{effective mubar EOMs}
\be
\dot{b}&=\f{1}{2\S(b)}\Big(\gamma^2\Lambda p_c-\big(\S(b)^2+\gamma^2\big)\Big)+\f{\delta_c}{\delta_b}\big(c\,\C(c)-\S(c)\big),\\
\dot{c}&=\f{\gamma^2\Lambda}{2}\f{p_b}{\S(b)}\left(1+\f{b\,\C(b)}{\S(b)}\right)-\S(c)-c\,\C(c)+\f{\delta_b}{2\delta_c}\left(b\,\C(b)\left(1-\f{\gamma^2}{\S(b)^2}\right)+\f{\gamma^2}{\S(b)}-\S(b)\right),\\
\dot{p}_b&=\f{p_b}{2\S(b)^2}\Big(\gamma^2\Lambda p_c+\big(\S(b)^2-\gamma^2\big)\Big)\C(b),\\
\dot{p}_c&=2p_c\,\C(c),
\ee
\end{subequations}
where we have introduced $\C(b)\coloneqq\cos(b\delta_b)$ which is such that $\C(b)\to1$ when $\Delta\to0$, and similarly for $c$. With this, it is straightforward to see that the limit $\Delta\to0$ leads back to the classical equations of motion \eqref{EOMs with standard N}. Note that the polymerized versions of the quantities \eqref{Dirac observables CD} are not conserved anymore, even in the case $\Lambda=0$, so we loose the explicit expression for the Dirac observables.

Unfortunately, the non-linearity of these effective equations prevents us from obtaining an exact solution, even in the case $\Lambda=0$, and we have to proceed numerically. In order to set the initial conditions for the numerical evolution, we impose that the effective solution approaches the classical solution in the regime of low curvature (i.e. near the horizon). Explicitly, we pick the initial conditions for $(b,p_b,p_c)$ at the point where the true classical solution \eqref{pb classical solution} for $p_b$ exhibits a maximum. As the effective constraint must be satisfied at every point of the solution, we then obtain $c(\tau_0)$ from the vanishing of the effective Hamiltonian constraint \eqref{effective Hamiltonian}. Note that in this case we do not have control over the effective integration constants (or Dirac observables), and the effective solutions are controlled by the choice of $(A,B)$ in \eqref{classical solutions with standard N}, or equivalently a chosen pair amongst $(M,C,D)$.

Plotting the solutions requires to choose a value for the cosmological constant. In reduced Planck units, its value is of order $10^{-122}$, which would effectively be treated as zero in the numerical evaluation of the solution. In unimodular gravity however, $\Lambda$ is a dynamical variable whose value can change between solutions, so it is not unreasonable to consider an arbitrary value. Since we are interested in the role of $\Lambda$ as a variable conjugated to a cosmological time to be used for the deparametrization of the quantum theory, we will restrict to the case $9M^2\Lambda<1$. In this regime, the physics of both the classical and effective solutions is qualitatively unchanged with respect to the case $\Lambda=0$. This will facilitate the comparison with the abundant results which exist in the literature for $\Lambda=0$. It is only in the quantum theory that $\Lambda\neq0$ will play a crucial role for us, since it will give us access to the preferred deparametrization. Taking $\Lambda=0$, the compared evolution of the true classical and effective trajectories is represented on figure \ref{fig1} below.

\begin{figure}[h]
$$\includegraphics[scale=0.6]{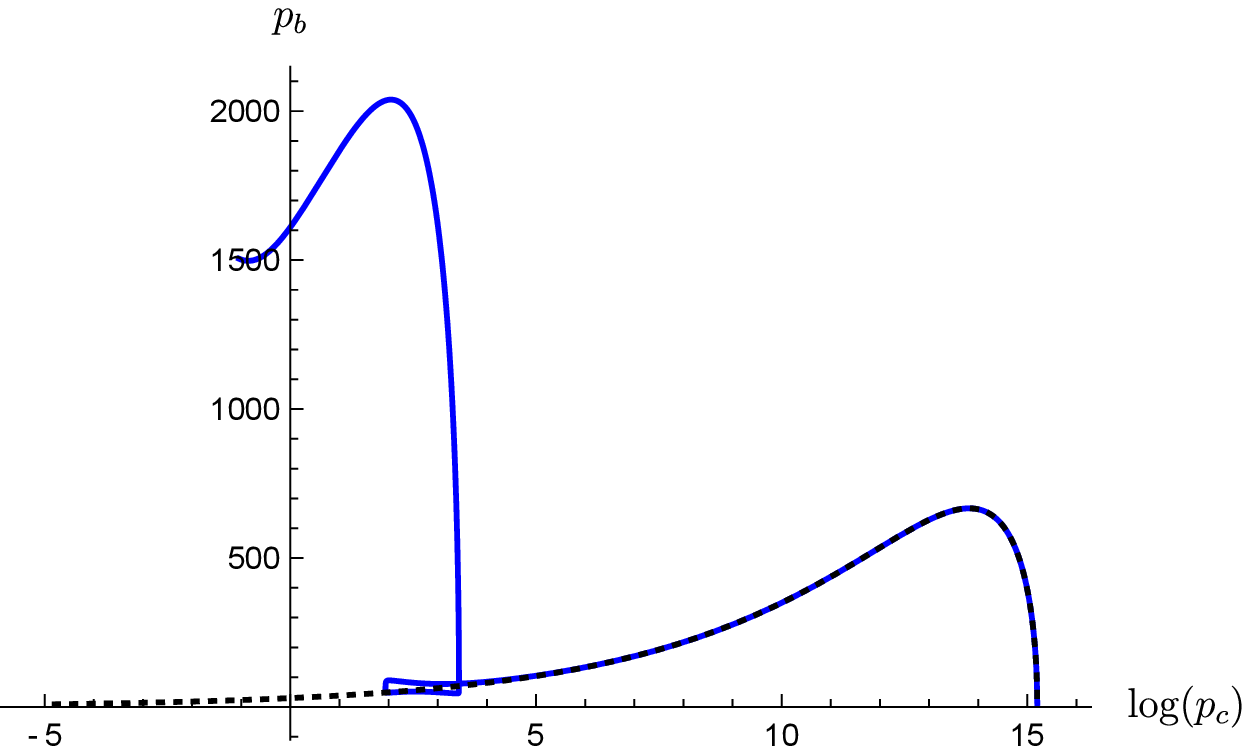}\q\includegraphics[scale=0.6]{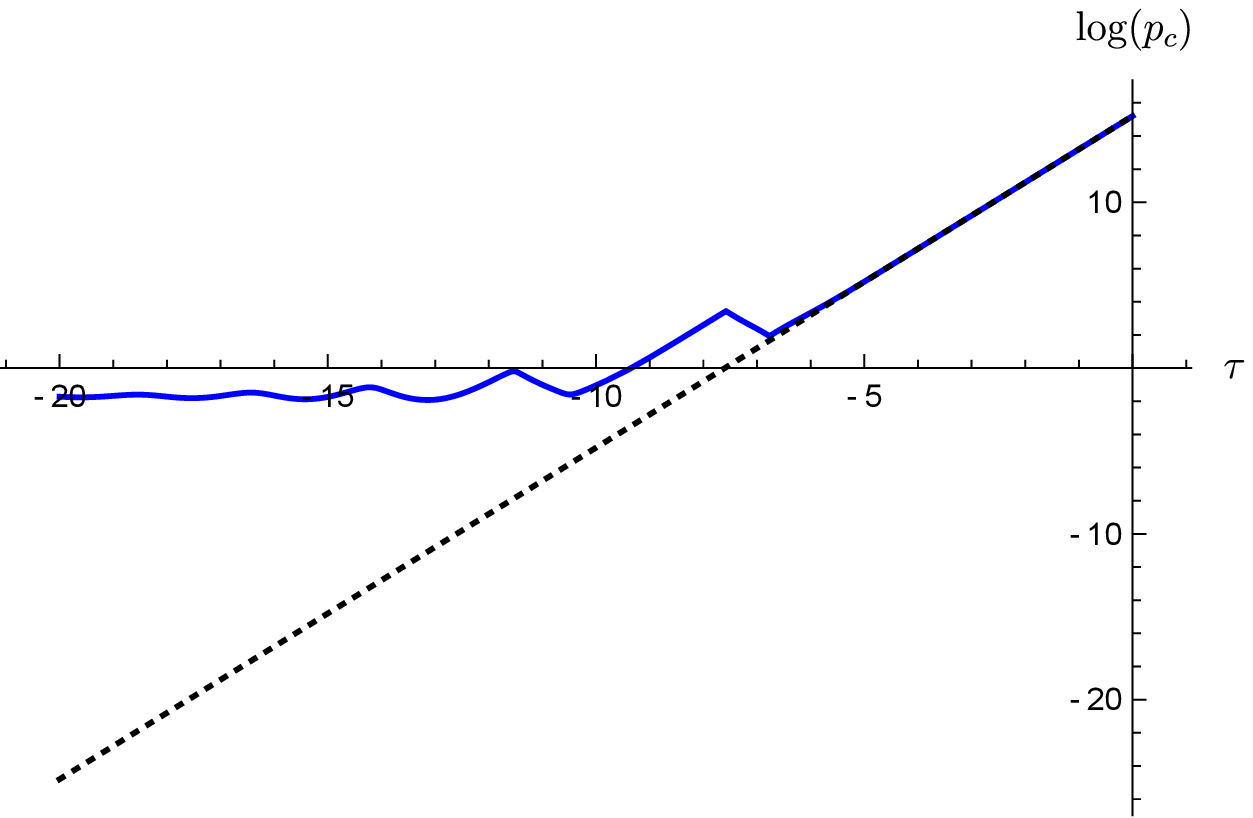}$$
\caption{Comparison of the classical (dashed) and effective (solid) evolution of $p_c$ in terms of $p_b$ (left) and time $\tau$ defined by the lapse \eqref{standard N} (right). For this plot we have set the initial conditions $A$ and $B$ in \eqref{classical solutions with standard N} by choosing a mass $M=10^4$ and $D=10^3$, and taken $\Lambda=0$.}\label{fig1}
\end{figure}

This plot puts in evidence the main issue with the $\bar{\mu}$ scheme applied to the black hole interior. One can see that $p_c$ initially follows the true decreasing classical trajectory, and then starts to oscillate in the Planck regime, undergoing several bounces with a decreasing trend on average. This however leads to an inconsistency, as $p_c$, which represents the radius of the 2-sphere, becomes at some point so small that the area of this 2-sphere is smaller than the area gap $\Delta$. As a consequence the plaquette $\square_{\theta\phi}$ can no longer fit on the 2-sphere. This violates the very construction of the $\bar{\mu}$ scheme, and therefore makes it inconsistent.

\subsection[$\mu_0$ scheme]{$\boldsymbol{\mu_0}$ scheme}
\label{sec:effective mu0}

One possibility in order to cure the issue encountered in the $\bar{\mu}$ scheme is to actually go back to constant $\delta$'s, as in the $\mu_0$ scheme \cite{Bohmer:2007wi,Corichi:2015xia}. In order to implement this, we require that the areas of the loops measured with the fiducial metric be equal to the area gap of LQG. This translates into the conditions $\text{Ar}_{x\theta}=L_0\delta_b\delta_c$ and $\text{Ar}_{\theta\phi}=\delta_b^2$, where once again the area of $\square_{\theta\phi}$ is effective since the corresponding loop is not closed. Solving for the $\delta$'s then gives
\be
\delta_b=\sqrt{\Delta},
\q\q
\delta_c=\f{\sqrt{\Delta}}{L_0}.
\ee
Again, this is compatible with the invariance under choice of $L_0$ since it fixes the product $L_0\delta_c$ to an invariant quantity.

With this choice the Hamiltonian is given again by \eqref{effective Hamiltonian}. The $\delta$'s being constants, the effective equations of motion now take the form.
\begin{subequations}
\be
\dot{b}&=\f{1}{2\S(b)}\Big(\gamma^2\Lambda p_c-\big(\S(b)^2+\gamma^2\big)\Big),\\
\dot{c}&=\gamma^2\Lambda\f{p_b}{\S(b)}-2\S(c),\\
\dot{p}_b&=\f{p_b}{2\S(b)^2}\Big(\gamma^2\Lambda p_c+\big(\S(b)^2-\gamma^2\big)\Big)\C(b),\\
\dot{p}_c &=2p_c\,\C(c),
\ee
\end{subequations}
which can be compared to \eqref{EOMs with standard N} and \eqref{effective mubar EOMs}. Once again, the polymerized versions of the quantities \eqref{Dirac observables CD} are not conserved anymore, so we loose the explicit expression for the Dirac observables.

In the case $\Lambda=0$ these effective equations of motion have an analytical solution since the equation for $c(\tau)$ decouples \cite{Bohmer:2007wi,Corichi:2015xia,Ashtekar:2018cay,Bodendorfer:2019jay}. Then, the polymerized version of $D$ given by\footnote{We use a tilde to denote the Dirac observables of the polymerized theory.} $\tilde{D}\coloneqq L_0^{-1}\S(c)p_c$ is a Dirac observable\footnote{Note that we need to include a factor of $L_0^{-1}$ in order to obtain the invariant combination $L_0\delta_c$ in the polymerized observable $L_0^{-1}\S(c)p_c=\sin(c\delta_c)p_c/(L_0\delta_c)$.} \cite{Ashtekar:2018cay,Bodendorfer:2019jay}, while the second one has a much more complicated expression \cite{Bodendorfer:2019jay}.

For $\Lambda\neq0$ we can solve these equations numerically by fixing the initial conditions in the classical regime as in the previous subsection. As mentioned above, one can check that the physics is qualitatively unaffected by the presence of the cosmological constant. At this point, it is interesting to discuss the physical meaning of the Dirac observables $C$ and $D$ in relation to the mass $M$ of the black hole. When using the classical solution \eqref{classical solutions with standard N} in order to fix the initial conditions, there are two natural choices for expressing the constants $A$ and $B$ in terms of the Dirac observables $C$ and $D$ and the mass. Indeed, one can write
\be\label{MCD initial condition}
A=\f{8M}{\sqrt{C}}=\f{4D}{3},
\q\q
B=4M^2,
\ee
which enables us to set initial conditions either on the pair $(M,C)$ or instead on the pair $(M,D)$. The figure below shows the plots of the classical and effective evolutions for initial conditions on $(M,D)$, where we have taken $\Lambda=10^{-8}$. Once again, the results are actually qualitatively insensitive to the choice of $\Lambda$ as long as $9M^2\Lambda<1$, which is indeed the regime of interest for us since then there is little to no departure from the well-studied case $\Lambda=0$.

\begin{figure}[h]
$$\includegraphics[scale=0.6]{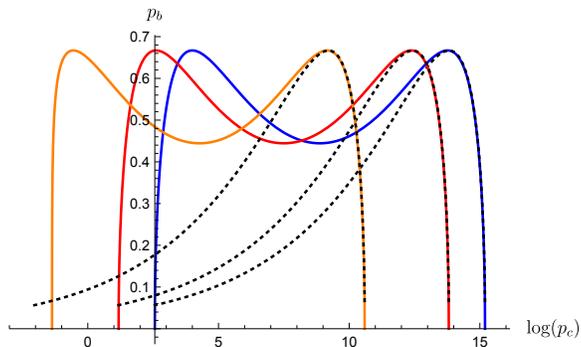}$$
\caption{Comparison of the classical (dashed) and effective (solid) evolution of $p_c$ in terms of $p_b$, for $D=1$ and various choices of initial black hole mass: $M=$ $10^3$ (blue), $5\times10^2$ (red), $10^2$ (orange). For the cosmological constant we have chosen $\Lambda=10^{-8}$.}\label{fig2}
\end{figure}

One can see on figure \ref{fig2} that for every trajectory the singularity is avoided by a bounce in the 2-sphere radius $p_c$. The effective quantum dynamics merges together  the initial black hole with a white hole solution. The transition surface is the 2-sphere where $p_c$ bounces. At this point, the expansion of the future-pointing null vector normal to the 2-sphere changes sign, indicating that the transition surface separates a trapped (black hole) from an anti-trapped (white hole) region. Since $\dot{p}_c$ has a single zero along each trajectory, each solution has a single transition surface.

The horizons are located at the points where $p_b=0$. The white hole horizon is where the evolution parameter $\tau$ reaches its lower bound $\tau_\text{min}$, while the black hole horizon is at $\tau_\text{max}$. The radius of the white hole horizon is then given by $\sqrt{p_c}(\tau_\text{min})$. When an explicit solution to the effective equations is available, as in the case $\Lambda=0$, one can explicitly reconstruct the effective line element and read off the radii of the black and white hole horizons. By doing so one relates these radii to the integration constants of the effective solutions, and the radii can then be used as independent boundary data that uniquely specify the solutions \cite{Bodendorfer:2019cyv,Bodendorfer:2019jay}. As expected, the effective black hole mass only agrees with the value $M$ which we have used to set the initial conditions up to small quantum corrections, which when $\Lambda\neq0$ can easily be evaluated numerically by comparing the classical and effective values of $\sqrt{p_c}(\tau_\text{max})$.

It is interesting to fix $C$ or $D$ and vary the black hole mass (i.e. using \eqref{MCD initial condition} to set the initial conditions) in order to study how the white hole mass depends on it. For each trajectory labelled by $M$ (at fixed $C$ or $D$), one simply compares the black hole radius $\sqrt{p_c}(\tau_\text{max})$ (classical or effective) with the while hole radius $\sqrt{p_c}(\tau_\text{min})$. This reveals a simple linear relationship $M_\text{WH}\propto M$, which can also be understood analytically with the thorough analysis of \cite{Bodendorfer:2019cyv}.

\subsection[Generalized $\mu_0$ scheme]{Generalized $\boldsymbol{\mu_0}$ schemes}

In the so-called generalized $\mu_0$ schemes, the polymerization parameters are allowed to depend on the mass $M$. As such, these approaches sit somehow in between the proper $\mu_0$ and $\bar{\mu}$ schemes since the mass is fixed for a given solution but can vary between different ones.

In \cite{Corichi:2015xia} for example, the authors have implicitly introduced a mass dependency in one of the polymerization parameters by taking
\be
\delta_b=\f{\sqrt{\Delta}}{M},
\q\q
\delta_c=\f{\sqrt{\Delta}}{L_0}.
\ee
With this choice, one can show that the there is again a bounce between a black hole and a white hole, but the mass of the white hole then scales as $M_\text{WH}\propto M^3$. This phenomenon has been referred-to as mass amplification.

In order to avoid this large mass amplification, the authors of \cite{Ashtekar:2018cay} have proposed an alternative criterion for fixing the polymerization parameters. This criterion is to fix the relationship between the physical area and the area gap at the level of the transition surface $\T$, i.e. to impose $2\pi\delta_b\delta_cp_b\big|_\T=\Delta$ and $4\pi\delta_b^2p_c\big|_\T=\Delta$. In the case $\Lambda=0$, the effective solutions enable to extract explicitly $p_c|_\T$, and to approximate $p_b|_\T$, and one finds \cite{Ashtekar:2018cay}
\be\label{AOS deltas}
\delta_b=\left(\f{\sqrt{\Delta}}{\sqrt{2\pi}\gamma^2M}\right)^{1/3},\q\q2L_0\delta_c=\left(\f{\gamma\Delta^2}{4\pi^2M}\right)^{1/3}.
\ee
When the explicit form of the effective solutions is not available, as in the case $\Lambda\neq0$, it is still possible to adapt this strategy. Indeed, now the solutions depend on the two initial conditions and on the values of $\delta_b$ and $\delta_c$. For a fixed value of the initial conditions and $\delta_b$, one can evaluate $\delta_b^2p_c\big|_\T$ for various values of $\delta_c$. Constraining this to equal $\Delta$ will then give a relationship between $\delta_c$ and $\delta_b$, which can then be inserted in $\delta_b\delta_cp_b\big|_\T=\Delta$ to find $\delta_b$. One can check that this reproduces accurately \eqref{AOS deltas}. With this prescription, one finds that the issue of mass amplification is cured, and the white hole mass scales as $M_\text{WH}\propto M$.

\subsection{Mixed scheme}
\label{sec:mixed scheme}

As we will see in the next section, in order to build the quantum theory it will be convenient to work with a different set of classical variables. Inspired by \cite{Bodendorfer:2019nvy,Bodendorfer:2019cyv}, we consider the canonical variables
\be\label{mixed variables}
p_1\coloneqq-\f{c}{2\gamma},
\q\q
v_1\coloneqq p_c,
\q\q
p_2\coloneqq\f{4}{\gamma}\f{b}{p_b},
\q\q
v_2\coloneqq-\f{p_b^2}{8},
\ee
which are such that
\be 
\{v_1,p_1\}=1=\{v_2,p_2\},
\ee 
with the other brackets vanishing. Note that this represents only half the change of variables of \cite{Bodendorfer:2019nvy,Bodendorfer:2019cyv}, as the pair $(p_1,v_1)$ is simply a rescaling of $(c,p_c)$, and only $(p_2,v_2)$ is a new canonical pair. This is the reason for which we keep lower case letters for $(p_1,p_2)$, at the difference with \cite{Bodendorfer:2019nvy,Bodendorfer:2019cyv}.

\subsubsection{Effective evolution}

With these new variables the classical Hamiltonian \eqref{H with standard N} corresponding to the choice of lapse \eqref{standard N} becomes
\be
\H=2p_1v_1+p_2v_2-2(1-\Lambda v_1)\f{1}{p_2}.
\ee
The classical dynamics in terms of these variables is given by
\begin{subequations}\label{mixed variables classical solutions}
\be
v_1(\tau)&=Be^{2(\tau-\tau_0)},\\
v_2(\tau)&=-\f{A^2}{8}\left(e^{(\tau-\tau_0)}-e^{2(\tau-\tau_0)}+\f{\Lambda}{3}Be^{4(\tau-\tau_0)}\right),\\
p_1(\tau)&=\f{A}{2}\left(\f{1}{2B}e^{-2(\tau-\tau_0)}-\f{\Lambda}{3}e^{(\tau-\tau_0)}\right),\\
p_2 (\tau)&=\f{4}{A}e^{-(\tau-\tau_0)},
\ee
\end{subequations}
and the Dirac observables become
\be\label{CD in terms of vp}
C=\f{16B}{A^2}=v_1p_2^2,
\q\q
D=\f{3A}{4}=p_1v_1-p_2v_2+\f{2}{p_2}.
\ee

In order to build the effective theory we are going to polymerize the variables $p_i$ by replacing them with
\be\label{mixed polymerization}
\S(p_i)\coloneqq\f{\sin(p_i\lambda_i)}{\lambda_i},
\ee
with $\lambda_i$ some polymerization parameters at Planck scale. Recall that in order for the classical theory to be independent of the fiducial length $L_0$, under a rescaling $L_0\to\alpha L_0$ the connection and triad variables must transform as $c\to\alpha c$ and $p_b\to\alpha p_b$. In terms of the new variables this gives
\be
v_1\to v_1,
\q\q
v_2\to\alpha^2v_2,
\q\q
p_1\to\alpha p_1,
\q\q
p_2\to\f{p_2}{\alpha},
\ee
which means that we have to pick the polymerization parameters such that $L_0\lambda_1$ and $L_0^{-1}\lambda_2$ are invariant under rescaling. In terms of dimensions, we must also have $[\lambda_i]=[p_i]^{-1}$. Since $[p_1]=\emptyset$ and $[p_2]=\text{length}^{-2}$, imposing a $\mu_0$ scheme leads to the prescription
\be
\lambda_1=\beta_1\f{\sqrt{\Delta}}{L_0},
\q\q
\lambda_2=\beta_2\sqrt{\Delta}L_0,
\ee
where $\beta_1$ and $\beta_2$ are possible dimensionless constants (which we will set to $\beta_1=1=\beta_2$ for the numerical analysis). In terms of the initial connection variables $(b,c)$, this choice is equivalent to choosing $\delta_b\propto1/p_b$ and $\delta_c=$ constant, which justifies the name of ``mixed'' scheme.

\begin{figure}[h]
$$\includegraphics[scale=0.6]{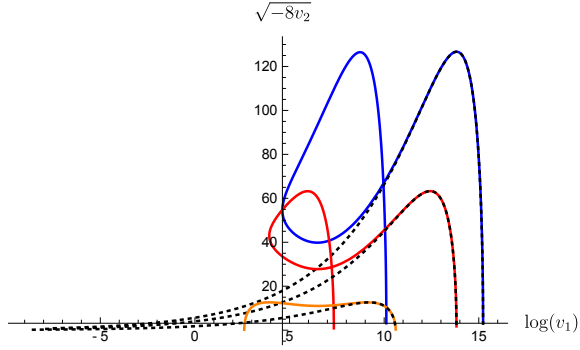}$$
\caption{Comparison of the classical (dashed) and effective (solid) evolution of $v_1$ in terms of $p_b=\sqrt{-8v_2}$, for $C=10^3$ and various choices of initial black hole mass: $M=$ $10^3$ (blue), $5\times10^2$ (red), $10^2$ (orange). For the cosmological constant we have chosen $\Lambda=10^{-8}$.}\label{fig3}
\end{figure}

From the polymerized Hamiltonian in this mixed scheme, we now get the following effective equations of motion:
\begin{subequations}
\be
\dot{v}_1 &=2v_1\,\C(p_1),\\
\dot{v}_2 &=v_2\,\C(p_2)+2(1-\Lambda v_1)\f{\C(p_2)}{\S(p_2)^2},\\
\dot{p}_1 &=-2\S(p_1)-\f{2\Lambda}{\S(p_2)},\\
\dot{p}_2 &=-\S(p_2),\\
\dot{T}&=\f{16\pi v_1}{\S(p_2)}.
\ee
\end{subequations}
Once again, these equations can unfortunately not be solved analytically in the case $\Lambda\neq0$. A numerical study reveals that the solutions are however qualitatively independent from $\Lambda$ in the regime of interest $9M^2\Lambda<1$. The numerical evolution for $\Lambda=10^{-8}$ is represented on figure \ref{fig3}. This shows once again a bounce in the two-sphere radius (represented by $v_1$) and a transition to a white hole.

One can also gather information about the effective solutions by solving the effective equations of motion in the case $\Lambda=0$. The analytic solutions in this case are given by
\begin{subequations}\label{mixed variables effective solutions no Lambda}
\be
v_1(\tau)&=\tilde{B}e^{2\tau}+\f{\tilde{A}^2\lambda_1^2}{64\tilde{B}}e^{-2\tau},\\
v_2(\tau)&=\Big(\tilde{A}^2(e^\tau-1)e^\tau+4\lambda_2^2\Big)\f{1}{2}\left(\f{1}{4}+\f{\lambda_2^2}{\tilde{A}^2}e^{-2\tau}\right),\\
p_1(\tau)&=\f{2}{\lambda_1}\arctan\left(\f{\tilde{A}\lambda_1}{8\tilde{B}}e^{-2\tau}\right),\\
p_2(\tau)&=\f{2}{\lambda_2}\arctan\left(\f{2\lambda_2}{\tilde{A}}e^{-\tau}\right),
\ee
\end{subequations}
where we have set the origin of the time parameter at $\tau_0=0$, and used a tilde to distinguish the integration constants of the effective solutions from that of the classical solutions\footnote{One could think that this is useless as these are integration constants, but we want to distinguish $(\tilde{A},\tilde{B})$ from $(A,B)$, which are related by $A=4D/3$ and $B=CD^2/9$ to the Dirac observables \eqref{CD in terms of vp} of the \textit{classial} theory.}. Consistently, one can see that the limit $(\lambda_1,\lambda_2)\rightarrow0$ of these solutions leads to the classical solutions \eqref{mixed variables classical solutions} with $\Lambda=0$. One can compare the exact effective solutions \eqref{mixed variables effective solutions no Lambda} with equations (3.26)--(3.29) of \cite{Bodendorfer:2019cyv} (or equivalently (3.11)--(3.14) of \cite{Bodendorfer:2019jay}), which are the effective solutions in variables $(v_1,v_2,P_1,P_2)$.

\subsubsection{Black and white hole radii and masses}

With the exact effective solutions \eqref{mixed variables effective solutions no Lambda} in the case $\Lambda=0$ at hand, we would like to discuss, following \cite{Bodendorfer:2019cyv,Bodendorfer:2019jay}, the relationship between the initial conditions $(\tilde{A},\tilde{B})$ and the back and white hole masses. For this, we must first reconstruct the effective interior line element.

Recalling that in the effective theory the variables $p_i$ are polymerized according to \eqref{mixed polymerization}, in terms of the variables \eqref{mixed variables} and with the polymerized lapse corresponding to \eqref{standard N}, the effective interior line element takes the form
\be
\de\tilde{s}^2=\f{2v_1\lambda_2^2}{v_2\sin^2(\lambda_2p_2)}\de\tau^2-\f{8v_2}{L_0^2v_1}\de y^2+v_1\de\Omega^2.
\ee
Redefining the coordinates as in the classical case, i.e. defining
\be
t\coloneqq\sqrt{\tilde{B}}e^\tau,
\q\q
x\coloneqq\f{\tilde{A}}{L_0\sqrt{\tilde{B}}}y,
\ee
we find the effective metric
\be
\de\tilde{s}^2=
\f{(t^2+t_+t_-)(t^4+a_0^2)}{(t-t_+)(t-t_-)t^4}\de t^2-\f{(t-t_+)(t-t_-)(t^2+t_+t_-)}{t^4+a_0^2}\de x^2+\left(t^2+\f{a_0^2}{t^2}\right)\de\Omega^2,
\ee
where
\be
t_\pm\coloneqq\f{\sqrt{\tilde{B}}}{2}\left(1\pm\sqrt{1-\f{16\lambda_2^2}{\tilde{A}^2}}\right),
\q\q
a_0\coloneqq\f{\tilde{A}\lambda_1}{8}.
\ee
One can see that the quantum effects, manifested here by the presence of the polymerization parameters $(\lambda_1,\lambda_2)$, have the effect of shifting the classical value $R_\text{BH}^\text{class}=t_+^\text{class}=\sqrt{B}$ of the black hole horizon radius, and of giving rise to a non-vanishing value of $t_-$. Recalling that the squared radius appears in front of the angular component of the metric and is given here by $v_1$, we obtain the black and white hole radii by evaluating
\begin{subequations}
\be
R_\text{BH}^2&=v_1(t_+)=t_+^2+\f{a_0^2}{t_+^2}=\tilde{B}+\f{\tilde{A}^2\lambda_1^2}{64\tilde{B}}+\circ\big(\lambda_2^2\big),\\
R_\text{WH}^2&=v_1(t_-)=t_-^2+\f{a_0^2}{t_-^2}=\f{\tilde{A}^6\lambda_1^2}{1024\tilde{B}\lambda_2^4}+\circ\left(\f{\lambda_1^2}{\lambda_2^2}\right).
\ee
\end{subequations}
Using this, one can choose to fix either $\tilde{A}$ or $\tilde{B}$ and obtain an (involved) expression for $R_\text{WH}$ as a function of $R_\text{BH}$.

Let us finally close this section by discussing, following \cite{Bodendorfer:2019cyv}, the ADM masses defined in the asymptotic regions. For this we consider the extension of the effective metric to the regions outside the horizons, where the time coordinate becomes space-like. In the limits $t\to\infty$ and $t\to0$ we find two asymptotically flat regions where one can define an ADM mass. In the first region the effective line element becomes
\be
\de\tilde{s}^2_+\approx-\left(\f{2M_\text{BH}}{t}-1\right)^{-1}\de t^2+\left(\f{2M_\text{BH}}{t}-1\right)\de x^2+t^2\de\Omega^2,
\q\q
M_\text{BH}=\f{\sqrt{\tilde{B}}}{2},
\ee
and the classical solution is therefore recovered (as expected far from the singularity). For the limit $t\to0$ we first need to define a new set of variables as
\be
t'\coloneqq\f{\tilde{A}\lambda_1}{8}\f{1}{t},
\q\q
x'\coloneqq\f{32\tilde{B}\lambda_2^2}{L_0\tilde{A}^3\lambda_1}y.
\ee
With this we then find
\be
\de\tilde{s}^2_-\approx-\left(\f{2M_\text{WH}}{t'}-1\right)^{-1}\de t'^2+\left(\f{2M_\text{WH}}{t'}-1\right)\de x'^2+t'^2\de\Omega^2,
\q\q
M_\text{WH}=\f{\tilde{A}^3\lambda_1}{64\sqrt{\tilde{B}}\lambda_2^2}.
\ee
Once again, one can see that this second region exists because of the quantum effects controlled by the polymerization parameters.

Now that the classical and effective dynamics in the variables $(v_1,v_2,p_1,p_2)$ has been studied, we can finally turn to the study of the quantization.

\section{Quantum dynamics}
\label{sec:quantum}

We now turn to the study of the quantum dynamics. Recall that our goal is to deparametrize the evolution with respect to the unimodular clock in order to obtain a true Schr\"odinger evolution equation. For this, we need to choose the lapse \eqref{unimodular N}, for which the classical Hamiltonian in $(p,v)$ variables becomes
\be\label{pv unimodular Hamiltonian}
\H=p_1p_2+\f{1}{v_1}\left(\f{1}{2}p_2^2v_2-1\right)+\Lambda.
\ee
Our task is now to regularize and represent the action of this Hamiltonian. We will see that the choice of $(p,v)$ variables enables to find analytical expressions for the eigenfunctions in the Wheeler--DeWitt (WDW) quantum theory. These can then be used to build the LQC quantum theory, following the construction of \cite{Ashtekar:2006uz,Ashtekar:2006wn}.

Note that the symmetry representing the orientation reversal of the triad must be considered. This means that the wavefunctions and the operators must be symmetric under $v_1\to-v_1$ (since $v_1=p_c$ and $v_2=-p_b^2/8$).

\subsection{Regularized Hamiltonian}

The first step in the construction of the quantum dynamics is to specify the Hilbert space and its basis states, and then to give the action of the Hamiltonian operator on these states.

Following the standard LQC construction, we choose the Hilbert space to be
\be\label{Hilbert spaces}
\H_\text{total}=\H_1\otimes\H_2,
\q\q
\H_i\coloneqq L^2(\R_\text{Bohr},\de{v_i}_\text{Bohr}),
\ee
where $\R_\text{Bohr}$ is the Bohr compactification of the real line \cite{Ashtekar:2003hd}, and then pick a basis of volume eigenstates $\ket{v_1,v_2}$ such that
\be
\hat{v}_1\ket{v_1,v_2}=v_1\ket{v_1 , v_2},
\q\q
\hat{v}_2\ket{v_1,v_2}=v_2\ket{v_1 , v_2}.
\ee
The conjugated operators $(\hat{p}_1,\hat{p}_2)$ are not represented on the Hilbert space $\H_\text{total}$, and only their exponentiated version exists. This fact is of course what has motivated the heuristic polymerization scheme discussed in the previous section. In the quantum theory we must therefore consider the regularized Hamiltonian constraint operator written in terms of the operators $\widehat{\S(p_i)}$ acting like\footnote{This follows simply from the action of exponentials of $p_i$ as translation operators.}
\begin{subequations}
\be
\widehat{\S(p_1)}\ket{v_1,v_2}&=\f{1}{2i\lambda_1}\big(\ket{v_1-\lambda_1,v_2}-\ket{v_1+\lambda_1,v_2}\big),\\
\widehat{\S(p_2)}\ket{v_1,v_2}&=\f{1}{2i\lambda_2}\big(\ket{v_1,v_2-\lambda_2}-\ket{v_1,v_2+\lambda_2}\big),
\ee
\end{subequations}
or in terms of wavefunctions
\begin{subequations}
\be
\widehat{\S(p_1)}\Psi(v_1,v_2)&=\f{1}{2i\lambda_1}\big(\Psi(v_1+\lambda_1,v_2)-\Psi(v_1-\lambda_1,v_2)\big),\\
\widehat{\S(p_2)}\Psi(v_1,v_2)&=\f{1}{2i\lambda_2}\big(\Psi(v_1,v_2+\lambda_2)-\Psi(v_1,v_2-\lambda_2)\big).
\ee
\end{subequations}

Now, we need to regularize the inverse power of $v_1$ appearing in the Hamiltonian, since the action of the corresponding naive operator on the $v_1=0$ eigenstates is ill-defined. For this we use the standard ``Thiemann trick'', which is based on the classical phase space identity
\be
\U(p_1)^{-1}\{|v_1|^n,\U(p_1)\}=in\lambda_1\text{sgn}(v_1)|v_1|^{n-1},
\ee
where $\U(p_1)\coloneqq e^{i\lambda_1p_1}$ and similarly for $p_2$. Once quantized, this identity enables us to define the regularized inverse volume operator in terms of the commutator involving a positive power of the volume. More precisely, taking $n=1/2$, transforming the classical identity into a quantum operator relation, and taking its square, we get the regularized operator\footnote{Note that we set $\hbar=1=G$. And we symmetrize the operator in order to keep the parity in $v_1$}
\be
\widehat{\f{1}{v_1}}=\f{1}{\lambda_1^2}\Big(\widehat{\U(p_1)^{-1}}\Big[\sqrt{|\hat{v}_1|}\ ,\ \widehat{\U(p_1)}\Big]-\widehat{\U(p_1)}\Big[\sqrt{|\hat{v}_1|}\ ,\ \widehat{\U(p_1)^{-1}}\Big]\Big)^2.
\ee
Its action on states is
\be
\widehat{\f{1}{v_1}}\ket{v_1,v_2}=\f{1}{\lambda_1^2}\big(\sqrt{|v_1-\lambda_1|}-\sqrt{|v_1+\lambda_1|}\big)^2\ket{v_1,v_2}\eqqcolon\B(v_1)\ket{v_1,v_2},
\ee
where we have defined the operator $\B(v_1)$ for later convenience. 

We now have all the ingredients to write the action on eigenfunctions of the quantum Hamiltonian constraint operator
\be
\widehat{\H}=\widehat{\S(p_1)}\,\widehat{\S(p_2)}+\widehat{\f{1}{v_1}}\left(\f{1}{2}\widehat{\S(p_2)}\,\hat{v}_2\,\widehat{\S(p_2)}-\mathds{1}\right)+\hat{\Lambda}.
\ee
Crucially, because we are considering unimodular gravity, the cosmological constant becomes the operator $-8\pi i\partial/\partial T$, and the action of the full Hamiltonian constraint becomes a Schr\"odinger equation of the form
\be\label{full pv Schrodinger}
-i\f{\partial\Psi(v_1,v_2)}{\partial T}=\Theta\Psi(v_1,v_2),
\ee
where we have defined the gravitational operator acting as
\be
&\f{\sgn(v_1)}{32\pi\lambda_1\lambda_2}\Big(\Psi(v_1+\lambda_1,v_2+\lambda_2)-\Psi(v_1-\lambda_1,v_2+\lambda_2)-\Psi(v_1+\lambda_1,v_2-\lambda_2)+\Psi(v_1-\lambda_1,v_2-\lambda_2)\Big)\cr
&+\f{\B(v_1)}{16\pi}\left(\f{1}{4\lambda_2^2}\Big[(v_2+\lambda_2)\Psi(v_1,v_2+2\lambda_2)-2v_2\Psi(v_1,v_2)+(v_2-\lambda_2)\Psi(v_1,v_2-2\lambda_2)\Big]+2\Psi(v_1,v_2)\right).\cr
&\eqqcolon\Theta\Psi(v_1,v_2).
\ee

Because of its structure, the Hamiltonian only relates the values of the wavefunction at some lattice points $v_i=\eps_i+n_i\lambda_i$, with $n_i\in\mathbb{Z}$ and $\eps_i\in[0,\lambda_i)$. This means that although the Hilbert spaces $\H_i$ in \eqref{Hilbert spaces} are non-separable, for each $\eps_i$ there exists a separable Hilbert subspace $\H_i^\eps\subset\H_i$ which is superselected and preserved under the evolution. In particular, we can fix a $\eps=0$ for simplicity, and focus on wavefunctions defined only on the corresponding lattice. The inner product is then given by
\be
\braket{\Psi|\Psi'}=\f{\lambda_1\lambda_2}{\pi^2}\sum_{v_i=n_i\lambda_i}\overline{\Psi}(v_1,v_2)\Psi'(v_1,v_2).
\ee
For the following discussion it will turn out to be more convenient to work in the $p_2$ representation, where the wavefunction is given by
\be
\Psi(v_1,p_2)\coloneqq\braket{v_1,p_2|\Psi}=\sum_{v_2}\braket{p_2|v_2}\Psi(v_1,v_2)=\sum_{v_2}e^{ip_2v_2}\Psi(v_1,v_2).
\ee 
As the Fourrier series of $\Psi(v_1,v_2)$, this wavefunction if periodic with period $2\pi/\lambda_2$. This is why only periodic functions of $p_2$ can be defined as multiplicative operators in this representation, while $\hat{p}_2$ is ill-defined. The functions $\S(p_2)$ and $\C(p_2)$ being well-defined, the Hamiltonian is acting in this representation as
\be\label{p2 rep operator}
\Theta\Psi(v_1,p_2)
&=i\f{\sgn(v_1)}{16\pi\lambda_1}\S(p_2)\Big(\Psi(v_1+\lambda_1,p_2)-\Psi(v_1-\lambda_1,p_2)\Big)\cr
&\phantom{=\ }-\f{B(v_1)}{16\pi}\left(i\S(p_2)\left[\C(p_2)\Psi(v_1,p_2)+\S(p_2)\f{\partial\Psi(v_1,p_2)}{\partial p_2}\right]-2\Psi(v_1,p_2)\right).
\ee
We can now study the WDW limit of this Hamiltonian and the associated quantum theory.

\subsection{Wheeler--DeWitt limit}

The WDW limit of the gravitational operator \eqref{p2 rep operator} is obtained in the limit $\lambda_i\to0$ for the polymerization parameters. In this case we obtain
\be\label{WDW operator}
\tilde{\Theta}\Psi(v_1,p_2)=i\f{\sgn(v_1)}{8\pi}p_2\f{\partial\Psi(v_1,p_2)}{\partial v_1}-\f{1}{16\pi|v_1|}\left(ip_2\left[\Psi(v_1,p_2)+p_2\f{\partial\Psi(v_1,p_2)}{\partial p_2}\right]-2\Psi(v_1,p_2)\right).
\ee
As expected, this is the operator we would have obtained from the Hamiltonian \eqref{pv unimodular Hamiltonian} by promoting the classical phase space variables to operators acting as
\be
\hat{v}_1\Psi=v_1\Psi,
\q\q
\hat{v}_2\Psi=i\f{\partial\Psi}{\partial p_2},
\q\q
\hat{p}_1\Psi=-i\f{\partial\Psi}{\partial v_1},
\q\q
\hat{p}_2\Psi=p_2\Psi,
\ee
on the wavefunctions $\Psi(v_1,p_2)$ in the Hilbert space $L^2(\R,\de v_1)\otimes L^2(\R,\de p_2)$.

\subsubsection{Dirac observables}

Our task is now to solve the Schr\"odinger equation $-i\partial_T\Psi(v_1,p_2)=\tilde{\Theta}\Psi(v_1,p_2)$. For this, we look for eigenfunctions of the 
gravitational operator such that
\be
\tilde{\Theta}\,\tilde{\mathbf{e}}_k(v_1,p_2)=k\,\tilde{\mathbf{e}}_k(v_1,p_2),
\ee
with $\Lambda=8\pi k$. The general solution is then
\be
\Psi(v_1,p_2,T)=\int\de k\,\tilde{\mathbf{e}}_k(v_1,p_2)\Psi(k)e^{ikT}.
\ee

Because we are dealing with two variables, these eigenfunctions will be degenerate. A possible way to lift this degeneracy is to find an operator commuting with the Hamiltonian, which can therefore be used to label the eigenstates. With our choice of variables, we can naturally define two quantum operators corresponding to the classical integrals of motion
\be
C=|v_1|p_2^2,
\q\q
D=\f{1}{2}(p_1v_1+v_1p_1-p_2v_2-v_2p_2)+\f{2}{p_2}.
\ee
The action of the corresponding operators on the wavefunctions is given by
\begin{subequations}
\be
\hat{C}\Psi(v_1,p_2)&=|v_1|p_2^2\Psi(v_1,p_2),\\
\hat{D}\Psi(v_1,p_2)&=-\left(i-\f{2}{p_2}+iv_1\f{\partial}{\partial v_1}+ip_2\f{\partial}{\partial p_2}\right)\Psi(v_1,p_2),
\ee
\end{subequations}
and one can easily check that these operators commute with the gravitational part of the Hamiltonian, i.e.
\be
[\tilde{\Theta},\hat{C}]=0=[\tilde{\Theta},\hat{D}].
\ee
On the other hand, the two observables do not commute with each other, but satisfy $[\hat{C},\hat D]=i\hat C$, which leads to an uncertainty relation that does not allow to specify both first integrals with arbitrary precision in the quantum theory.

We can now look for the eigenfunctions in the $(v_1,p_2)$ representation.

\subsubsection{Eigenfunctions}

Using $D$ to label the energy eigenvalues, we can lift the degeneracy and we find that
\be
\tilde{\mathbf{e}}_{k,D}(v_1,p_2)=\sqrt{\f{2}{9\pi}}\f{1}{p_2}\exp\left(\f{2i}{3p_2}\big(3-8\pi k|v_1|\big)\right)\exp\left(\f{iD}{3}\log\big(|v_1|p_2^2\big)\right)
\ee
is the solution to
\be
\tilde{\Theta}\,\tilde{\mathbf{e}}_{k,D}(v_1,p_2)=k\,\tilde{\mathbf{e}}_{k,D}(v_1,p_2),
\q\q
\hat{D}\,\tilde{\mathbf{e}}_{k,D}(v_1,p_2)=D\,\tilde{\mathbf{e}}_{k,D}(v_1,p_2),
\ee
with
\be
\braket{\tilde{\mathbf{e}}_{k,D}|\tilde{\mathbf{e}}_{k',D'}}=\int\de v_1\,\de p_2\,\overline{\tilde{\mathbf{e}}}_{k,D}(v_1,p_2)\,\tilde{\mathbf{e}}_{k',D'}(v_1,p_2)=\delta(k-k')\delta(D-D').
\ee
Remark that in order to perform this integral it is useful to change the variables to $z\coloneqq|v_1|/p_2$ and $w\coloneqq\log(|v_1|p_2^2)$. In fact, the WDW theory is most easily formulated in terms of these variables, but this has the disadvantage that that polymerization procedure is then cumbersome. Finally, note that an analytic expression for the eigenfunctions also exists with the $(v_1,v_2,P_1,P_2)$ variables defined in \cite{Bodendorfer:2019nvy,Bodendorfer:2019cyv}, and that the reason for which we have chosen to work instead with $(p_1,p_2)$ is that these variables are adapted to the unimodular clock.

With the above eigenfunctions, the general solution to the Schr\"odinger evolution equation is given by
\be
\Psi(v_1,p_2,T)=\int\de k\,\de D\,\tilde{\mathbf{e}}_{k,D}(v_1,p_2)\Psi(k,D)e^{ikT}.
\ee
We can now create a semi-classical state peaked around a classical solution with cosmological constant $\Lambda^*=8\pi k^*$ and first integral of motion $D^*$. For this, we take
\be
\Psi(k,D)=\f{1}{\sqrt{\pi\sigma_k\sigma_D}}\exp\left(-\f{(k-k^*)^2}{2\sigma_k^2}\right)\exp\left(-\f{(D-D^*)^2}{2\sigma_D^2}\right),
\ee
which gives
\be
\Psi(v_1,p_2,T)=\f{\N}{p_2}\exp\left[-\left(\f{T}{8\pi}-\f{2}{3}z\right)^2\f{\sigma_k^2}{2}-\f{\sigma_D^2}{2}w^2\right]\exp\left[ik^*\left(\f{T}{8\pi}-\f{2}{3}z\right)+\f{2i}{p_2}+\f{iD^*}{2}w\right],
\ee
where $z\coloneqq|v_1|/p_2$ and $w\coloneqq\log(|v_1|p_2^2)$. It turns out that this solution is also peaked on a particular value of the other first integral of motion, as we have
\be
\braket{\hat{C}}_\Psi=\braket{|v_1|p_2^2}_\Psi=1.
\ee
It is possible to change this value to any $C^*$ by simply rescaling the function as
\be
\Psi(k,D)\to\Psi(k,D)\exp\left(-\f{iD}{3}\log C^*\right).
\ee
Finally, at a given cosmological time we have
\be
\left\langle\f{2}{3}\f{|v_1|}{p_2}\right\rangle_\Psi=\f{T}{8\pi}=\f{2}{3}\f{|v_1|}{p_2}\Bigg|_\text{class}.
\ee
This shows that we have constructed semi-classical states of the WDW quantum theory peaked around a classical solution. These states can now be used to investigate the loop quantization of the model.

\subsection{Loop quantization}

We now come back to the study of the LQC quantum dynamics. There, the eigenfunctions will be found by using the knowledge of the WDW theory at late times, and evolved with the Schr\"odinger evolution through the (avoided) singularity.

\subsubsection{Eigenfunctions}

Let us now consider the LQC gravitational operator $\Theta$ acting in the $(v_1,p_2)$ representation as \eqref{p2 rep operator}. We look for eigenfunctions such that
\be\label{LQC eigenfunction}
\Theta\,\mathbf{e}_k(v_1,p_2)=k\,\mathbf{e}_k(v_1,p_2).
\ee
As in the WDW case treated above, the presence of two variables leads to an infinite degeneracy of the eigenfunctions for a given energy eigenvalue. Here unfortunately, as discussed in the section about the effective solutions, we do not have access to a polymerized Dirac observable which can be used to label the eigenstates, as we did using $D$ in the WDW case. Nevertheless, it is reasonable to expect that in the WDW limit $|v_1|\to\infty$ and $p_2\to0$ (or $\lambda_2\rightarrow0$) the eigenfunctions we are looking for satisfy
\be
\mathbf{e}_k(v_1,p_2)\xrightarrow[\substack{|v_1|\to\infty\\p_2\to0}]{}\int\de D'\,f(D')\,\tilde{\mathbf{e}}_{k,D'}(v_1,p_2),
\ee
for some functional $f(D)$. It is therefore natural to \textit{define} the eigenfunctions $\mathbf{e}_{k,D}(v_1,p_2)$ with the choice $f(D')=\delta(D-D')$.

Following the construction of \cite{Ashtekar:2006uz,Ashtekar:2006wn}, we can now try to match the values of $\mathbf{e}_{k,D}(v_1,p_2)$ with that of the WDW eigenfunctions at early times (i.e. in the low curvature regime), and then use the time-independent Schr\"odinger equation to evaluate the wavefunction. However, a complication in doing so is that the operator $\Theta$ acting in \ref{LQC eigenfunction} is discrete in the variable $v_1$ while continuous in $p_2$, meaning that we could only approximate the result numerically. Fortunately, it is possible to use an alternative approximation, which is to take \textit{only} the limit $v_1\to\infty$ in the action of $\Theta$ on the left-hand side of \ref{LQC eigenfunction}. Doing so, it is still possible to find an analytical solution for the eigenfunctions. First, let us use the limit of large $v_1$ to define the gravitational operator
\be
\underline{\Theta}\Psi(v_1,p_2)
&\coloneqq i\f{\sgn(v_1)}{8\pi}\S(p_2)\f{\partial\Psi(v_1,p_2)}{\partial v_1}\cr
&\phantom{\coloneqq\ }-\f{1}{16\pi|v_1|}\left(i\S(p_2)\left[\C(p_2)\Psi(v_1,p_2)+\S(p_2)\f{\partial\Psi(v_1,p_2)}{\partial p_2}\right]-2\Psi(v_1,p_2)\right).\q
\ee
One can see that this is the polymerized (in $p_2$) version of the WDW operator \eqref{WDW operator}. The eigenfunctions are then given by
\be\label{LQC approx eigen}
\mathbf{\underline{e}}_{\,k,D}(v_1,p_2)
&=\sqrt{\f{2}{9\pi}}\f{1}{\S(p_2)}\exp\left(\f{2i}{3\S(p_2)}\big(3\C(p_2)-8\pi k|v_1|\big[2-\C(p_2)\big]\big)\right)\exp\left(\f{iD}{3}\log\left[\f{4|v_1|\S(p_2)^2}{\big(\C(p_2)+1\big)^2}\right]\right),
\ee
and one can check that they indeed satisfy the properties
\be
\underline{\Theta}\,\underline{\mathbf{e}}_{\,k,D}(v_1,p_2)=k\,\underline{\mathbf{e}}_{\,k,D}(v_1,p_2),
\q\q
\underline{\mathbf{e}}_{\,k,D}(v_1,p_2)\xrightarrow[p_2\to0]{}\tilde{\mathbf{e}}_{k,D}(v_1,p_2).
\ee
We can now match the values of $\mathbf{e}_{k,D}$ and $\underline{\mathbf{e}}_{\,k,D}$ at some point $v_1^*\gg\lambda_1$ for all the values of $p_2$, and use equation \eqref{LQC eigenfunction} to find its values on the $v_1$ lattice. The eigenfunctions obey the following limits:
\be
\mathbf{e}_{k,D}(v_1,p_2)\xrightarrow[|v_1|\to\infty]{}\underline{\mathbf{e}}_{\,k,D}(v_1,p_2)\xrightarrow[p_2\to0]{}\tilde{\mathbf{e}}_{k,D}(v_1,p_2).
\ee
Note that we still find $\mathbf{e}_{k,D}(v_1,p_2)$ with a numerical approximation since the derivative term $\partial\Psi/\partial p_2$ is evaluated numerically (with the exception of the first step when $v_1=v_1^*$, where we have its analytical expression). Finally let us remark that because of the parity symmetry these eigenfunction must be symmetrized. The plots of the eigenfunctions are given below.

\begin{figure}[H]
$$\includegraphics[scale=0.6]{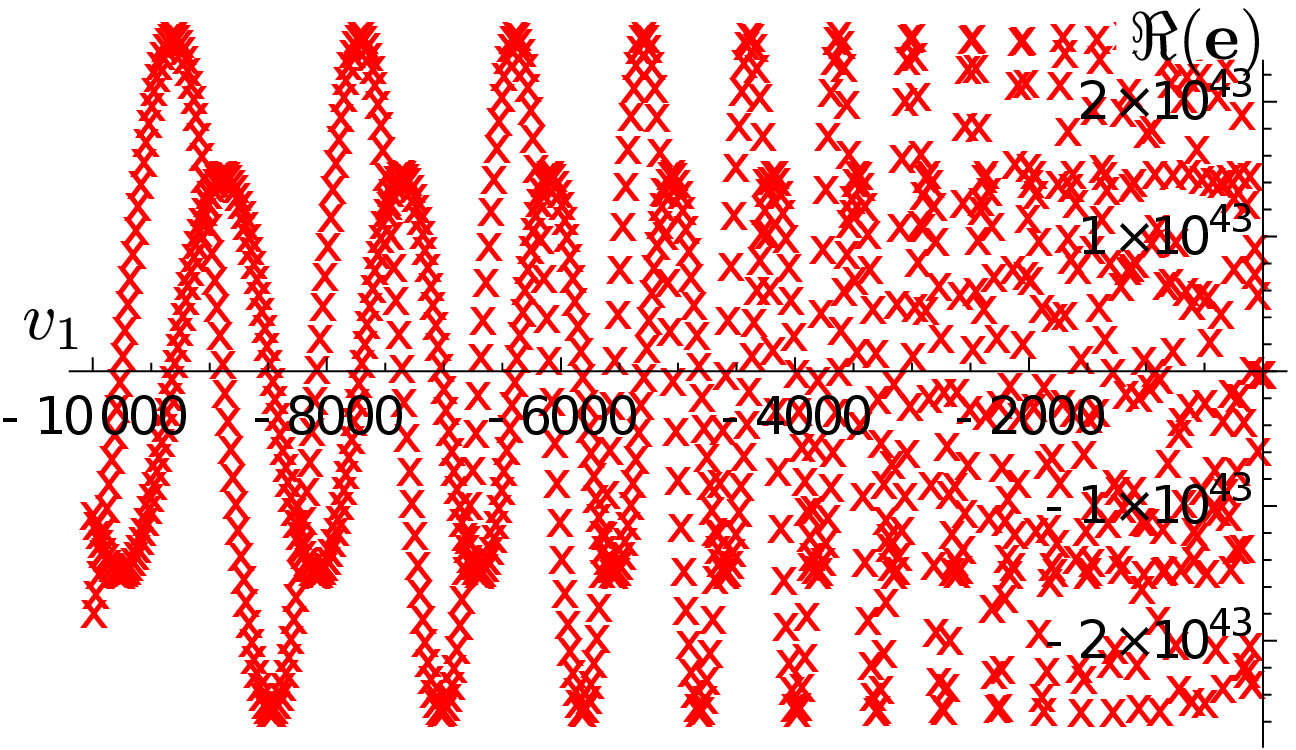}\q\includegraphics[scale=0.6]{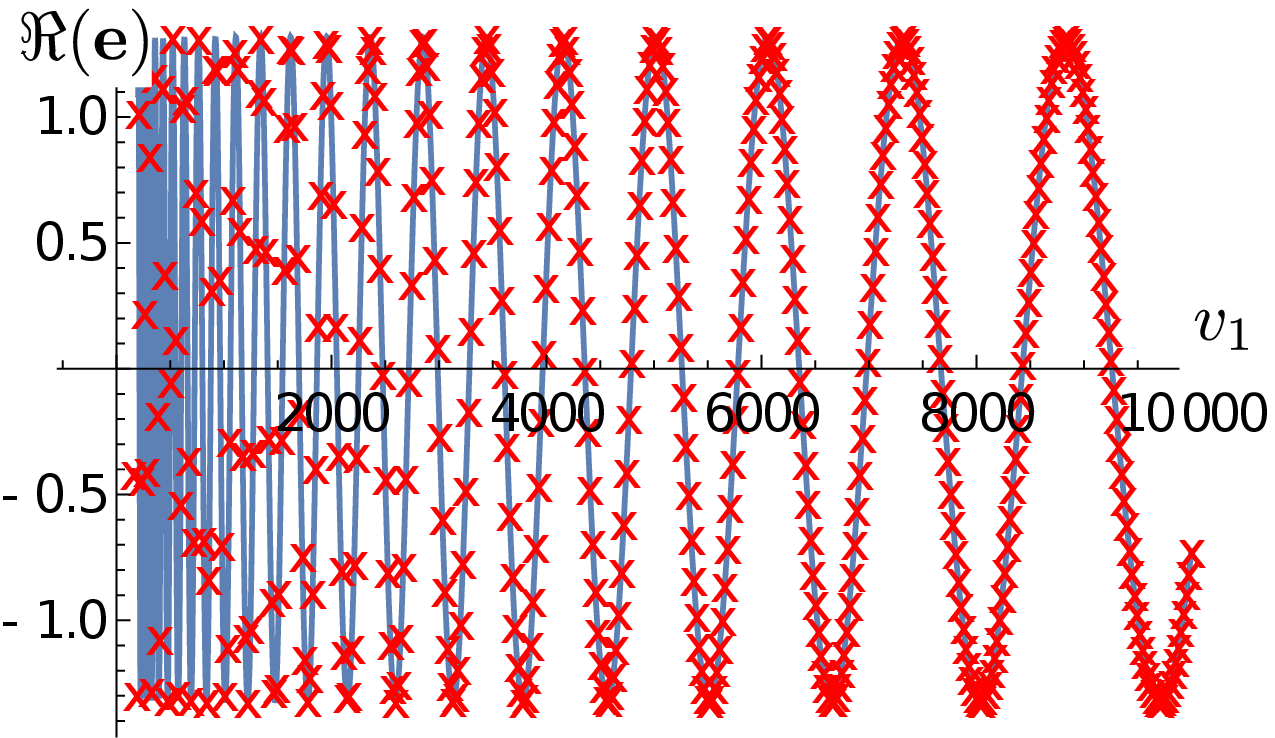}$$
\caption{Plot of (the real part of) an eigenfunction $\mathbf{e}_{k,D}(v_1,p_2)$ (before symmetrization) as a function of $v_1$ at fixed $p_2$. Here we take $p_2=\pi/(10\lambda_2)$. On the left the amplified zone is shown for $v_1\to-\infty$, and on the right we see that it is well approximated for $v_1\to+\infty$ by the solid blue line representing the WDW eigenfunctions $\underline{\mathbf{e}}_{\,k,D}(v_1,p_2)$. The numerical values used for these plots are $D=100$ and $\Lambda=10^{-8}$.}
\end{figure}

\begin{figure}[H]
$$\includegraphics[scale=0.6]{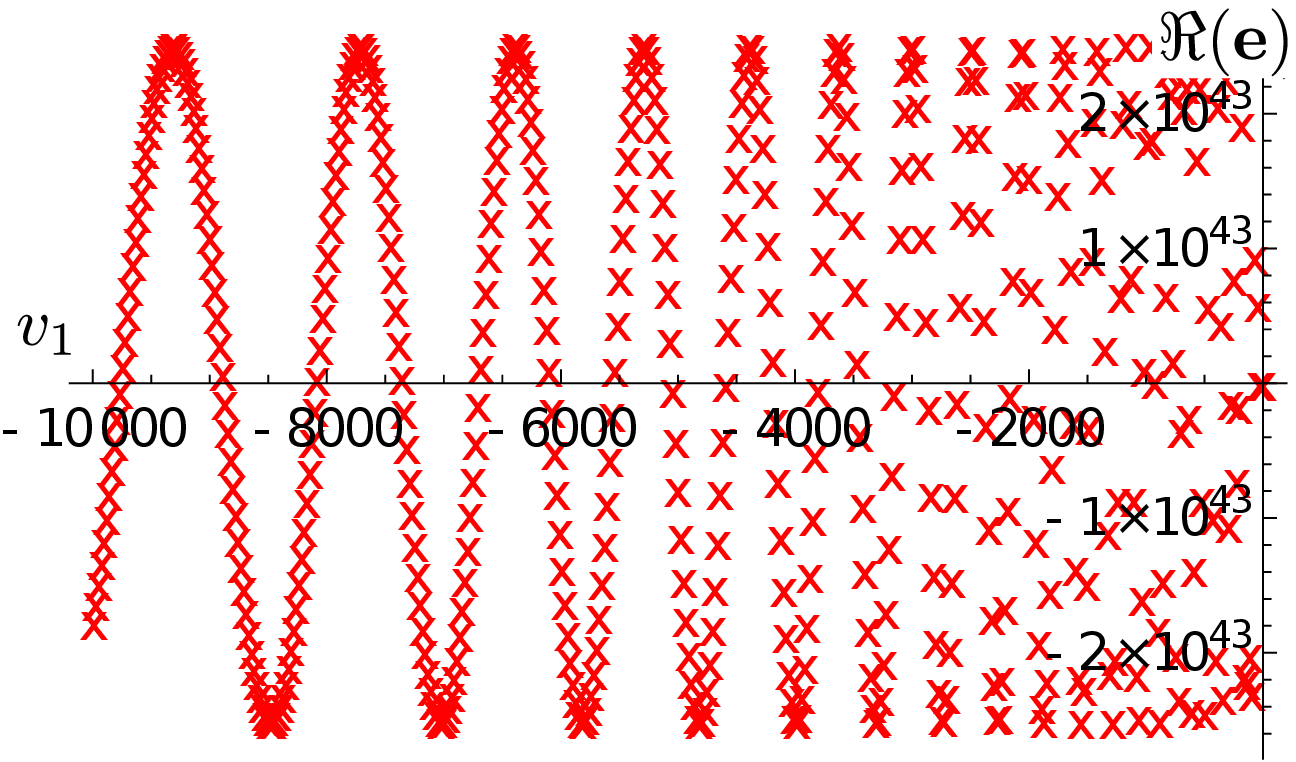}\q\includegraphics[scale=0.6]{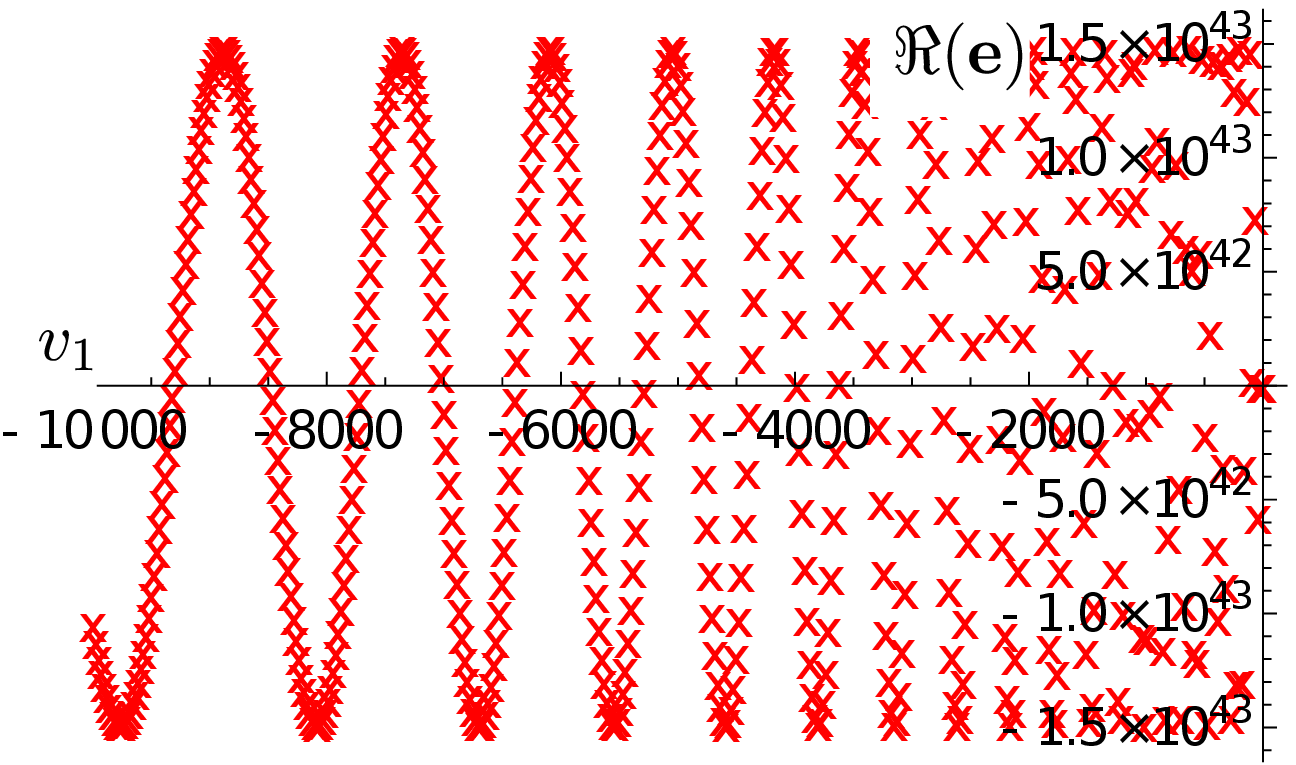}$$
\caption{Separation of the two outgoing waves, with supports on the lattices $v_1^*-2n\lambda_1$ on the left, and $v_1^*-(2n+1)\lambda_1$ on the right, for $n\to+\infty$.}
\end{figure}

We see that the eigenstates obtained here for the black hole interior in the loop quantization scheme exhibit the same features as that of the FLRW LQC eigenstates \cite{Ashtekar:2006uz,Ashtekar:2006wn}. More precisely, they are very well approximated by the WDW eigenstates in the classical regime away from the singularity, and they oscillate for $v_1\to\pm\infty$. This behavior can be separated as a ``superposition'' of two waves with support on even and odd lattices.

We note that a particular case is given by the eigenfunctions corresponding to a zero eigenvalue of the energy (i.e without cosmological constant). In this case, following the procedure described in the previous section, it is possible to get rid of the derivative term $\partial\Psi/\partial p_2$, and thus find a recurrence series in $v_1$ independent of $p_2$. Indeed, if $\Lambda=0$ in \eqref{LQC approx eigen} we see that the eigenfunction can be written as the product of two functions which depend on only one of the two variables, i.e.
\be
\mathbf{\underline{e}}_{\,0,D}(v_1,p_2)=\sqrt{\f{2}{9\pi}}\f{1}{\S(p_2)}\exp\left(\f{2i\C(p_2)}{\S(p_2)}+\f{iD}{3}\log\left[\f{4\S(p_2)^2}{\big(1+ \C(p_2)\big)^2}\right]\right)\exp\left(\f{i D}{3}\log|v_1|\right).
\ee
This, combined with \eqref{LQC eigenfunction} and the fact that at some point $\mathbf{e}_{\,0,D}(v_1^*,p_2)=\underline{\mathbf{e}}_{\,0,D}(v_1^*,p_2)$, gives the recursion relation
\be
\mathbf{e}_{k,D}(v_1-\lambda_1,p_2)=\mathbf{e}_{k,D}(v_1+\lambda_1,p_2)-2\sgn(v_1)B(v_1)\lambda_1\f{iD}{3}\mathbf{e}_{k,D}(v_1,p_2).
\ee
This would still require to be evaluated numerically however.

\subsubsection{Quantum evolution}

We can now consider the time-dependent Schr\"odinger equation \eqref{full pv Schrodinger} and use it to let evolve a wavepacket $\Psi(v_1,v_2)$ with respect to the cosmological time. This is to be contrasted with what is usually done in e.g. FLRW LQC \cite{Ashtekar:2006uz,Ashtekar:2006wn}, where one uses a massless scalar field as internal time and obtains in the deparametrized theory a Klein--Gordon evolution equation. With the unimodular clock the evolution can unambiguously be computed. For this, as initial condition $\Psi|_{T_0}$ we take a Gaussian semiclassical state peaked around a classical solution in a low-curvature regime for a given mass. More precisely, we take the same initial condition as for the effective evolution (i.e. when $v_2$ has a minimum) and write
\be
\Psi|_{T_0}(v_1, v_2)=\N\exp\left(-\f{(|v_1|-v_1^*)^2}{4\sigma_1^2}\right)\exp\left(-\f{(|v_2|-v_2^*)^2}{4\sigma_2^2}\right)\exp\Big(ip_1^*(|v_1|-v_1^*)+ip_2^*(|v_2|-v_2^*)\Big),
\ee
where $v_i^*=|v_i(T_0)|\big|_\text{class}$ and $p_i^*=p_i(T_0)\big|_\text{class}$, for the solution of mass $M$ and integration constant $C=1$. The data is then evolved with the Schr\"odinger equation using a fourth order Runge--Kutta method (RK4). At each time step we calculate the expectation values of the $v_i$ to compare them with the effective classical evolution of section \ref{sec:mixed scheme}. We need to restrict the domain of integration to a set $|v_i|<N_i\lambda_i$ with $N_i\gg1$. The boundary of the domain is chosen to be sufficiently far form the peak of the initial state by demanding that, at the boundary, the value of the wavefunction is less than $10^{-10}$ times the value at the peak. Moreover, we approximate the difference terms in equation \eqref{full pv Schrodinger} \textit{only at the boundary} by
\be
\Psi(v_i+\lambda_i)-\Psi(v_i-\lambda_i)\simeq\f{1}{2}\big(\Psi(v_i)-\Psi(v_i-\sgn(v_i)\lambda_i)\big).
\ee 

\begin{figure}[h]
$$\includegraphics[scale=0.5]{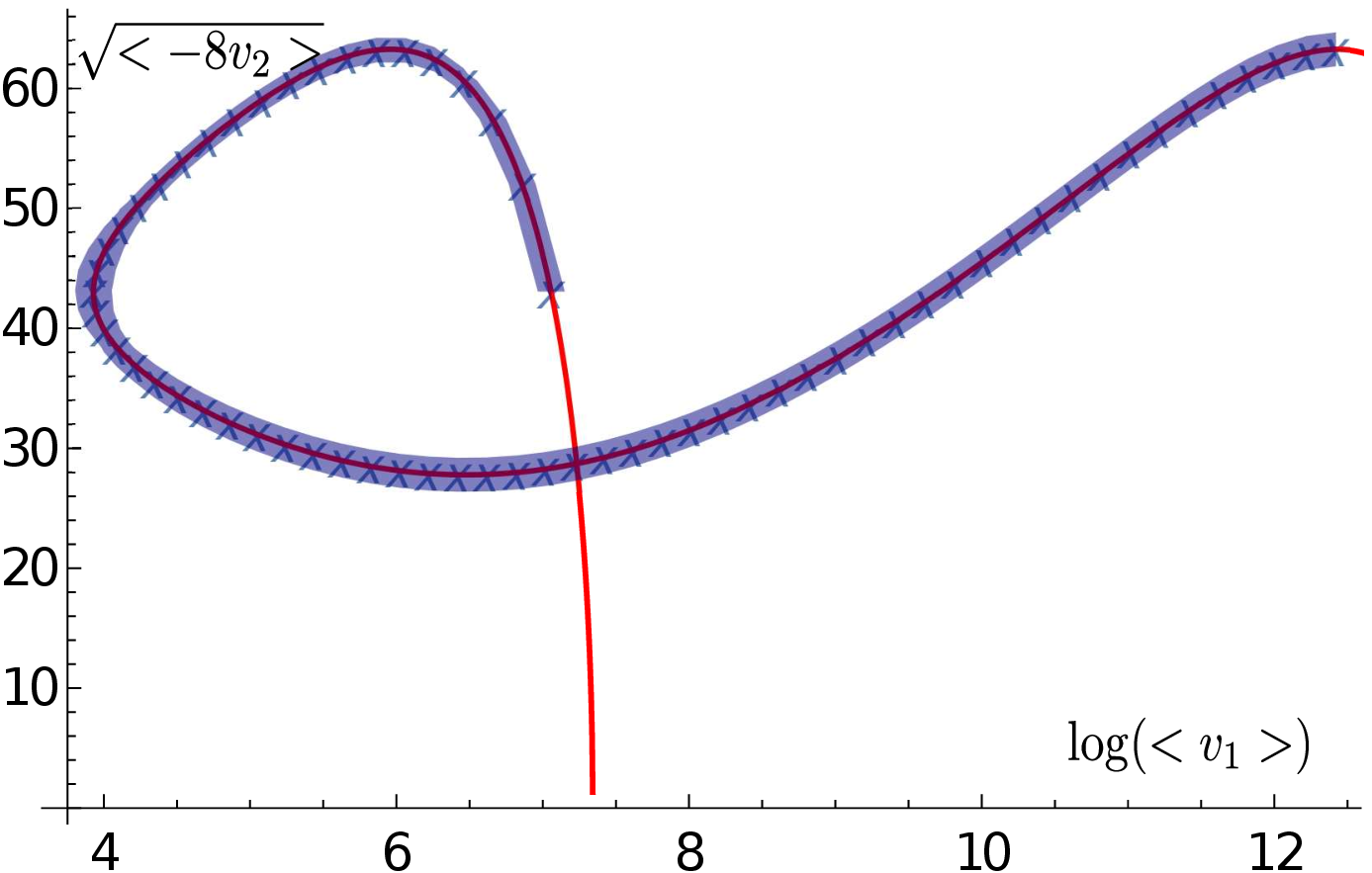}\q\includegraphics[scale=0.5]{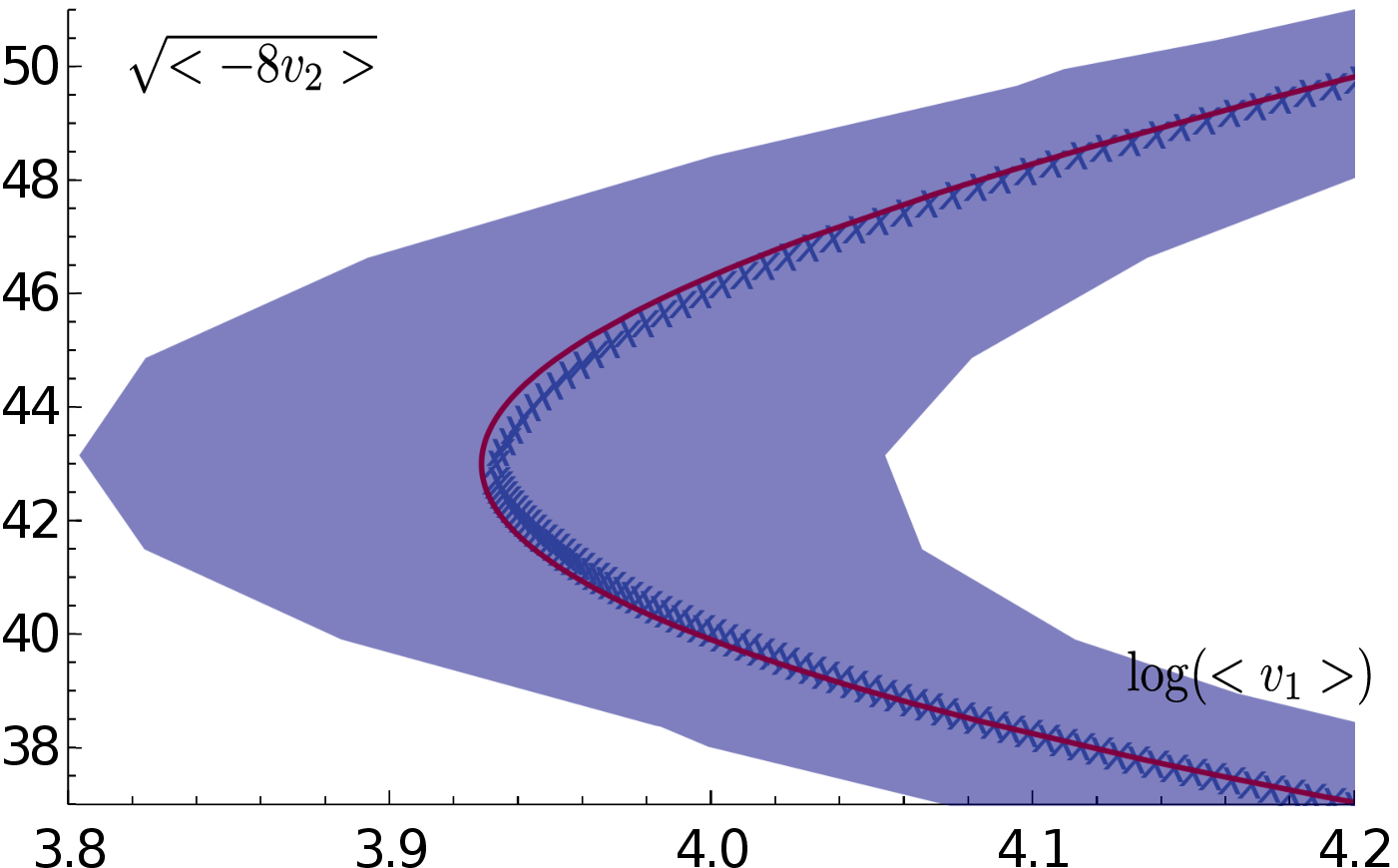}$$
\caption{Expectation values (and dispersions at $\pm\sigma$) of the wavefunction throughout the quantum evolution, compared to the effective evolution in red. The numerical values are $M=500$ and $C=10^3$. The plot on the right is zoomed around the bounce in $v_1$.}
\end{figure}

We see that the state remains sharply peaked throughout the evolution, and undergoes a quantum bounce.The expectation values of the wavefunction are very well approximated by the effective trajectory. Provided one has decided on a choice of variables, a regularization scheme, and found late time WDW eigenfunctions to evolve, this illustrates how the use of the unimodular clock enables to simply obtain the full quantum evolution.

With this first illustration of the usefulness of the unimodular representation for quantum cosmological models beyond flat FLRW, one can now envision studying other regularization schemes, evolving different quantum states, and performing a detailed numerical analysis of the quantum theory. This can hopefully serve as an investigation tool for polymerized models of the black hole interior, and be used to ask other physical questions such as that of unitarity of the evolution across the bounce \cite{Amadei:2019ssp,Amadei:2019wjp}.

\section{Perspectives}

In this paper we have presented the construction of the quantum evolution of the black hole interior spacetime using the framework of unimodular gravity. This is a simple reformulation of general relativity which is equivalent to Einstein's theory at the classical level, but which has the advantage of solving the problem of time in mini-superspace quantum cosmology. This is done by promoting the Hamiltonian constraint to a true Hamiltonian generating evolution along a ``cosmological time'', which as we have recalled in appendix \ref{appendix:unimodular} is just a measure of the elapsed four-volume between hypersurfaces. In quantum cosmology, this cosmological time reduces the study of the quantum dynamics to that of a Schr\"odinger evolution of Gaussian wavepackets peaked on semi-classical states at late times and labelled by Dirac observables. Following \cite{Ashtekar:2006uz,Ashtekar:2006wn}, we have computed this evolution by matching the wavefunctions to the WDW ones in the late time semi-classical regime. When evolved with the regularized LQC gravitational operator, these states remain sharply peaked and go through the bounce while following closely the effective classical trajectory.

These preliminary results show that the use of unimodular gravity for the study of the quantum dynamics of the black hole interior reproduces the qualitative features of the effective classical evolution. The use of the unimodular clock provides two important improvements: $i)$ the possibility to apply the framework to other regularization schemes for the Hamiltonian constraint, and to compare the details of the quantum dynamics in order to discriminate between the various schemes proposed e.g. in \cite{Ashtekar:2018cay,Bodendorfer:2019cyv,Bodendorfer:2019nvy,Bodendorfer:2019jay}; and $ii)$ the possibility of studying quantum dynamics of other mini-superspace models without having to define the square root of the (possibly negative definite) gravitational Hamiltonian in order to obtain a Schr\"odinger evolution equation for the states. This requires to extend the present construction to other regularization schemes, and to perform a detailed study of the numerics. In addition, it would be interesting to extend the construction of \cite{Amadei:2019ssp,Amadei:2019wjp} in order to study the Hawking information puzzle in the black hole interior spacetime. We note that, even disregarding the issues of regularizations of the Hamiltonian constraint arising in LQC, the unimodular representation can prove useful in order to investigate WDW mini-superspace quantum cosmology and black holes.

\section*{Acknowledgement}

We would like to thank Alejandro Perez and Lautaro Amadei for sharing their enthusiasm on unimodular gravity.

\appendix

\section{Unimodular gravity and cosmological time}
\label{appendix:unimodular}

Unimodular gravity was initially introduced by Einstein Hilbert in order to solve the ``cosmological constant problem''. The initial observation is that imposing the so-called unimodular condition $\det(g_{\mu\nu})=-1$ in the variation of the action leads to the trace-free Einstein field equations, from which one can obtain the full set of field equations provided we introduce the cosmological constant as a simple integration constant. We refer interested reader to the online resource \cite{unimodular_web}, which lists formulations of unimodular gravity and applications in various contexts. 

Here we will only focus on the generally-covariant formulation proposed in \cite{Henneaux:1989zc}, which illustrates very simply the interplay between the unimodular condition and the presence of a cosmological time variable. In this formulation, the cosmological constant is introduced as a Lagrange multiplier enforcing the unimodular condition. The gravitational action takes the form
\be\label{HT unimodular action}
S=\f{1}{16\pi}\int_M\de^4x\,\Big(\sqrt{-g}\,(R-2\Lambda)+2\Lambda\partial_\mu\tau^\mu\Big),
\ee
where $\tau^\mu=(\tau^0,\tau^a)$ is a space-time vector density. Performing a variation with respect to the multiplier $\Lambda$ gives the unimodular condition
\be\label{unimodular}
\sqrt{-g}=\partial_\mu\tau^\mu,
\ee
while varying $\tau^\mu$ leads to $\partial_\mu\Lambda=0$, which indicates that $\Lambda$ is a space-time constant that can be identified with the cosmological constant.

In order to interpret the vector density $\tau^\mu$, suppose that the manifold $M$ is diffeomorphic to $\Sigma\times\mathbb{R}$, with $\Sigma$ a compact slice. Integrating \eqref{unimodular} over the space-time region $R\subset M$ enclosed between two -hypersurfaces $\Sigma(t_1)$ and $\Sigma(t_2)$ then gives
\be
\int_R\de^4x\,\partial_\mu\tau^\mu=\int_R\de^4x\,\sqrt{-g}=\text{four volume enclosed between $\Sigma(t_1)$ and $\Sigma(t_2)$}.
\ee
Clearly, the canonical structure of the action \eqref{HT unimodular action} reveals that the cosmological constant $\Lambda$ is conjugated to the dynamical variable
\be
T(t)\coloneqq\int_{\Sigma(t)}\de^3x\,\tau^0.
\ee
This is a function which increases continuously along any future directed time-like curve, and which satisfies
\be
T(t_2)-T(t_1)=\text{four volume enclosed between $\Sigma(t_1)$ and $\Sigma(t_2)$}.
\ee
The evolution of the vector density $\tau^\mu$ is therefore pure gauge, apart from the component $\tau^0$, which gives rise to the unimodular time variable.

\section{Densitized triad, Ashtekar--Barbero connection, and Hamiltonian}
\label{appendix:EAH}

Following the literature on Kantowski--Sachs space-times in LQC, we focus here on line elements of the form
\be\label{KS metric}
\de s^2=-N^2\de t^2+\f{p_b^2}{L_0^2|p_c|}\de x^2+|p_c|\de\Omega^2.
\ee
We will furthermore consider $p_c\geq0$. With this metric the Einstein--Hilbert action becomes
\be\label{EH action}
\f{1}{16\pi}\int_M\de^4x\,\sqrt{-g}\,(R-2\Lambda)=\f{1}{16\pi L_0}\int_M\de^4x\,\sin\theta\left(\f{2Np_b}{\sqrt{p_c}}+\f{p_c'(p_bp_c'-4p_b'p_c)}{2Np_c^{3/2}}-2N\Lambda p_b\sqrt{p_c}\right).
\ee
In LQG we work instead with densitized triads and the canonically conjugated Ashtekar--Barbero connection, in terms of which the action takes the form
\be\label{AB action}
\int_M\de^4x\left(\f{1}{8\pi\gamma}E^a_i(A^i_a)'-NH-N^aH_a-\lambda^iG_i\right),
\ee
where $(H,H_a,G_i)$ are the scalar, vector, and Gauss constraints. Only the former is non-trivial with our choice of homogeneous metric. Its smeared form is the Hamiltonian
\be
\H\coloneqq\int_\Sigma\de^3x\,NH=\f{1}{16\pi}\int_\Sigma\de^3x\left(\f{N}{\sqrt{q}}E^a_iE^b_j\Big({\eps^{ij}}_kF^k_{ab}-2(1+\gamma^2)K^i_{[a}K^j_{b]}\Big)+2N\Lambda\sqrt{q}\right).
\ee
Here $q$ is the determinant of the spatial metric $q_{ab}$, anti-symmetrization of indices is defined with a factor 1/2, $\gamma$ is the Barbero--Immirzi parameter, and the curvature of the connection is given by $F^i_{ab}=\partial_aA^i_b-\partial_bA^i_a+{\eps^i}_{jk}A^j_aA^k_b$.

Let us now construct the various quantities entering the definition of the Hamiltonian and symplectic structure. The metric \eqref{KS metric} has non-vanishing Christoffel coefficients given by
\begin{subequations}
\begin{align}
\Gamma^t_{tt}&=\f{N'}{N},
&\Gamma^t_{xx}&=\f{p_b(2p_b'p_c-p_bp_c')}{2L_0^2N^2p_c^2},
&\Gamma^t_{\theta\theta}&=\f{p_c'}{2N^2},\\
\Gamma^t_{\phi\phi}&=\f{p_c'}{2N^2}\sin^2\theta,
&\Gamma^x_{xt}&=\f{2p_b'p_c-p_bp_c'}{2p_bp_c},
&\Gamma^\theta_{t\theta}&=\f{p_c'}{2p_c},\\
\Gamma^\theta_{\phi\phi}&=-\cos\theta\,\sin\theta,
&\Gamma^\phi_{\phi t}&=\f{p_c'}{2p_c},
&\Gamma^\phi_{\phi\theta}&=\cot\theta,
\end{align}
\end{subequations}
where the prime denotes derivative with respect to $t$. The tetrad coefficients $e^I_\mu$ are related to the metric by $g_{\mu\nu}=e^I_\mu e^J_\nu\eta_{IJ}$, where the internal Lorentz metric is $\eta_{IJ}=\text{diag}(-1,1,1,1)$. A possible choice (up to internal Lorentz gauge transformations) of tetrad coefficients for the metric \eqref{KS metric} is
\be\label{KS tetrad}
e^0_\mu=N\partial_\mu t,\q\q e^1_\mu=\f{p_b}{L_0\sqrt{p_c}}\partial_\mu x,\q\q e^2_\mu=\sqrt{p_c}\,\partial_\mu\theta,\q\q e^3_\mu=\sqrt{p_c}\sin\theta\,\partial_\mu\phi.
\ee
From these coefficients, we can then compute the coefficients $E^a_i\coloneqq\eps^{abc}\eps_{ijk}e^j_be^k_c/2$ of the densitized triad $E=E^a_i\tau^i\partial_a$. The non-vanishing components are
\begin{subequations}
\be
E^x_1&=e^2_\theta e^3_\phi-e^2_\phi e^3_\theta=p_c\sin\theta,\\
E^\theta_2&=e^3_\phi e^1_x-e^3_x e^1_\phi=\f{p_b}{L_0}\sin\theta,\\
E^\phi_3&=e^1_xe^2_\theta-e^1_\theta e^2_x=\f{p_b}{L_0},
\ee
\end{subequations}
which implies that we have
\be\label{KS E}
E=E^a_i\tau^i\partial_a=p_c\sin\theta\,\tau^1\partial_x+\f{p_b}{L_0}\sin\theta\,\tau^2\partial_\theta+\f{p_b}{L_0}\tau^3\partial_\phi.
\ee
We can now solve the torsion equation
\be
\partial_\mu e^I_\nu-\Gamma^\sigma_{\mu\nu}e^I_\sigma+\omega^I_{\mu J}e^J_\nu=0
\ee
in order to find the Lorentz connection coefficients compatible with the tetrad \eqref{KS tetrad}. This gives
\begin{subequations}
\begin{align}
\omega^{01}_\mu&=\f{2p_b'p_c-p_bp_c'}{2L_0Np_c^{3/2}}\partial_\mu x,
&\omega^{12}_\mu&=0,\\
\omega^{02}_\mu&=\f{p_c'}{2N\sqrt{p_c}}\partial_\mu\theta,
&\omega^{13}_\mu&=0,\\
\omega^{03}_\mu&=\f{p_c'}{2N\sqrt{p_c}}\sin\theta\,\partial_\mu\phi,
&\omega^{23}_\mu&=-\cos\theta\,\partial_\mu\phi.
\end{align}
\end{subequations}
With these connection coefficients we can then introduce the Ashtekar--Barbero connection with components
\be
A^i_\mu\coloneqq\Gamma^i_\mu+\gamma\omega^{0i}_\mu=-\f{1}{2}{\eps^i}_{jk}\omega^{jk}_\mu+\gamma\omega^{0i}_\mu.
\ee
We find
\begin{subequations}
\begin{align}
A^1_\mu=-\omega^{23}_\mu+\gamma\omega^{01}_\mu&=\cos\theta\,\partial_\mu\phi+\gamma\f{2p_b'p_c-p_bp_c'}{2L_0Np_c^{3/2}}\partial_\mu x,\\
A^2_\mu=-\omega^{31}_\mu+\gamma\omega^{02}_\mu&=\gamma\f{p_c'}{2N\sqrt{p_c}}\partial_\mu\theta,\\
A^3_\mu=-\omega^{12}_\mu+\gamma\omega^{03}_\mu&=\gamma \f{p_c'}{2N\sqrt{p_c}}\sin\theta\,\partial_\mu\phi.
\end{align}
\end{subequations}
Since $A$ and $E$ are canonically conjugated, denoting the momenta of the triad variables by $b$, and $c$, we can finally write
\be\label{KS A}
A=A^i_a\tau_i\de x^a=\f{c}{L_0}\tau_1\de x+b\tau_2\de\theta+b\sin\theta\,\tau_3\de\phi+\cos\theta\,\tau_1\de\phi,
\ee
where
\be
b\coloneqq\gamma\f{p_c'}{2N\sqrt{p_c}},\q\q c\coloneqq\gamma\f{2p_b'p_c-p_bp_c'}{2Np_c^{3/2}}.
\ee
With these identifications we can indeed verify that the canonical terms in the two actions \eqref{EH action} and \eqref{AB action} coincide. In the Hamiltonian formulation we are going of course to treat $(b,c)$ and $(p_b,p_c)$ as independent variables.

We can now use the expressions \eqref{KS E} and \eqref{KS A} to evaluate the Hamiltonian and the symplectic structure in connection-triad variables. For the Hamiltonian we can first compute the pieces
\be
E^a_iE^b_j{\eps^{ij}}_kF^k_{ab}
&=E^a_iE^b_j(2{\eps^{ij}}_k\partial_aA^k_b+A^i_aA^j_b-A^j_aA^i_b)\cr
&=2E^\theta_2E^\phi_3\partial_\theta A^1_\phi+2\big(E^x_1E^\theta_2A^1_xA^2_\theta+E^x_1E^\phi_3A^1_xA^3_\phi+E^\theta_2E^\phi_3A^2_\theta A^3_\phi\big)\cr
&=2\sin^2\theta\,p_b^2+2\big(E^x_1E^\theta_2A^1_xA^2_\theta+E^x_1E^\phi_3A^1_xA^3_\phi+E^\theta_2E^\phi_3A^2_\theta A^3_\phi\big)\cr
&=\f{2}{L_0^2}\sin^2\theta\,p_b\big(2bcp_c+(b^2-1)p_b\big),
\ee
and
\be
2E^a_iE^b_jK^i_{[a}K^j_{b]}
&=E^a_iE^b_j(K^i_aK^j_b-K^j_aK^i_b)\cr
&=2\big(E^x_1E^\theta_2K^1_xK^2_\theta+E^x_1E^\phi_3K^1_xK^3_\phi+E^\theta_2E^\phi_3K^2_\theta K^3_\phi\big)\cr
&=\f{2}{\gamma^2}\big(E^x_1E^\theta_2A^1_xA^2_\theta+E^x_1E^\phi_3A^1_xA^3_\phi+E^\theta_2E^\phi_3A^2_\theta A^3_\phi\big)\cr
&=\f{2}{\gamma^2L_0^2}\sin^2\theta\,p_b(2bcp_c+b^2p_b),
\ee
where we have used the fact that $\gamma K^i_a=A^i_a-\Gamma^i_a$ and only $\Gamma^1_\phi\neq0$. We then have to compute the integrals over $\Sigma$ in the Hamiltonian and the symplectic structure. However, the integrals of the homogeneous fields on $\Sigma$ are over $x\in\mathbb{R}$, $\theta\in[0,\pi]$, and $\phi\in[0,2\pi]$, and therefore divergent. This can be dealt with by restricting the integration to a fiducial finite interval $x\in[0,L_0]$. With this, and using that
\be
\sqrt{q}=\f{p_b\sqrt{p_c}}{L_0}\sin\theta,
\ee
one can finally show that the Hamiltonian reduces to expression \eqref{Hamiltonian} given in the main text (where the factors of $L_0$ have dropped out), and that the symplectic structure becomes
\be
\f{1}{8\pi\gamma}\int_\Sigma\de^3x\,E^a_i(A^i_a)'=\f{b'p_b}{\gamma}+\f{c'p_c}{2\gamma},
\ee
which implies the Poisson brackets \eqref{PBs}.

Note that it is only the ratios $L_0^{-1}p_b$ and $L_0^{-1}c$ which have an invariant meaning under the rescaling $L_0\to\alpha L_0$ of the fiducial interval introduced for the regularization of the spatial integrals.

\section{Classical dynamics with unimodular clock}
\label{appendix:unimodular N}

Here we study the classical dynamics with respect to the unimodular time variable. This is achieved by choosing the lapse
\be\label{unimodular N}
N=\f{2}{p_b\sqrt{p_c}}.
\ee
With this choice the Hamiltonian \eqref{Hamiltonian} becomes
\be\label{H with unimodular N}
\H=-\f{1}{\gamma^2p_bp_c}\Big(2bcp_c+(b^2+\gamma^2)p_b\Big)+\Lambda,
\ee
and as expected the gravitational part is proportional to the cosmological constant. The equations of motion are then found to be
\begin{subequations}\label{EOMs with unimodular N}
\be
\dot{b}&=\f{2}{\gamma}\f{bc}{p_b^2},\\
\dot{c}&=\f{2}{\gamma}\f{b^2+\gamma^2}{p_c^2},\\
\dot{p}_b&=\f{2}{\gamma}\left(\f{c}{p_b}+\f{b}{p_c}\right),\\
\dot{p}_c&=\f{4}{\gamma}\f{b}{p_b},\\
\dot{T}&=8\pi,\\
\dot{\Lambda}&=0,
\ee
\end{subequations}
while the vanishing of the Hamiltonian gives again the (lapse-independent) relation \eqref{c on constraint surface}. The solution to the equations of motion is
\begin{subequations}\label{classical solutions with unimodular N}
\be
b(\tau)&=\pm\gamma\big(g(\tau-\tau_0)\big)^{1/2},\\
c(\tau)&=\f{4\gamma}{3}\left(\f{\Lambda}{\sqrt{C}}\left(\f{3\sqrt{C}}{2}(\tau-\tau_0)\right)^{1/3}\!-D\left(3\sqrt{2C}(\tau-\tau_0)\right)^{-2/3}\right),\\
p_b(\tau)&=\pm2\left(\f{12}{C}(\tau-\tau_0)\right)^{1/3}\big(g(\tau-\tau_0)\big)^{1/2},\\
p_c(\tau)&=\left(\f{3\sqrt{C}}{2}(\tau-\tau_0)\right)^{2/3},\\
T(\tau)&=8\pi(\tau-\tau_0)-T_0,\label{T solution}
\ee
\end{subequations}
where we have introduced
\be
g(\tau)\coloneqq\f{D\sqrt{C}}{3}\left(\f{3\sqrt{C}}{2}\tau\right)^{-1/3}+\f{\Lambda}{3}\left(\f{3\sqrt{C}}{2}\tau\right)^{2/3}-1.
\ee
Equation \eqref{T solution} confirms the fact that $T$ is indeed a time variable since it evolves linearly in the coordinate time $\tau$. Here we have parametrized the solutions with $C$ and $D$ given in \eqref{Dirac observables CD}, which remain Dirac observables even with the choice of lapse\footnote{Recall that it is not guaranteed that Dirac observables with respect to a given Hamiltonian, i.e. a choice of lapse, still remain Dirac observables with respect to another choice of lapse. For this the Dirac observables have to commute with $\H_1/\H_2$, which is the ``relative'' lapse between the two Hamiltonians.} \eqref{unimodular N}. 

With the solutions to the classical equations of motion at hand, one can go back to the parametrization \eqref{homogeneous metric}, and write this line element as
\be
\de s^2=-g(\tau)^{-1}\left(\f{\sqrt{C}}{18}\f{1}{\tau^2}\right)^{2/3}\de\tau^2+g(\tau)\f{16}{L_0^2C}\de y^2+\left(\f{3\sqrt{C}}{2}\tau\right)^{2/3}\de\Omega^2.
\ee
Redefining new coordinates via
\be
\tau=\f{2}{3\sqrt{C}}t^3,\q\q y=\f{L_0\sqrt{C}}{4}x,
\ee
one gets that
\be
g(\tau)=-f(t)=-\left(1-\f{2M}{t}-\f{\Lambda}{3}t^2\right),
\ee
and we find the homogeneous interior line element \eqref{KdS} with a mass
\be
M=\f{D\sqrt{C}}{6},
\ee
in agreement with the result of section \ref{sec:classical}.

\bibliographystyle{Biblio}
\bibliography{Biblio.bib}

\end{document}